\documentclass[11pt]{elsarticle}
\usepackage[utf8]{inputenc}
\usepackage{url}

\usepackage{lineno}

\usepackage[a4paper, total={6in, 8in}]{geometry}

\usepackage{etoolbox}
\makeatletter

\patchcmd{\ps@pprintTitle}
  {\let\@oddhead\@empty}
  {\def\@oddhead{\mbox{}\hfill\thepage}}
  {}{}
\makeatother

\usepackage{psfrag,epsfig,graphics}
\usepackage{amsmath,amsthm,amssymb,multirow}
\usepackage{mathbbol}
\usepackage{amssymb}             

\DeclareSymbolFontAlphabet{\amsmathbb}{AMSb}%
\usepackage{mathabx} 

\usepackage{booktabs}
\usepackage{graphicx,amsmath,graphicx}
\usepackage{caption}
\usepackage[position=b]{subcaption}

\usepackage{multirow}
\usepackage[lined,linesnumbered,ruled]{algorithm2e}

\usepackage{color}  
\def\cred{\textcolor{red}}

\def\cmag{\textcolor{magenta}}

\newcommand{\cb}[1]{{\boldsymbol{#1}}}
\newcommand{\cp}[1]{\ifmmode {\mathcal{#1}}\else ${\mathcal{#1}}$\fi}

\newcommand{\bC}{\boldsymbol{C}}
\newcommand{\bD}{\boldsymbol{D}}

\newcommand{\bF}{\boldsymbol{F}}
\newcommand{\bG}{\boldsymbol{G}}
\newcommand{\bH}{\boldsymbol{H}}
\newcommand{\bI}{\boldsymbol{I}}
\newcommand{\bK}{\boldsymbol{K}}

\newcommand{\bP}{\boldsymbol{P}}
\newcommand{\bQ}{\boldsymbol{Q}}
\newcommand{\bR}{\boldsymbol{R}}
\newcommand{\bS}{\boldsymbol{S}}

\newcommand{\bT}{\boldsymbol{T}}

\newcommand{\bc}{\boldsymbol{c}}

\newcommand{\bh}{\boldsymbol{h}}

\newcommand{\bq}{\boldsymbol{q}}
\newcommand{\br}{\boldsymbol{r}}
\newcommand{\by}{\boldsymbol{y}}
\newcommand{\bs}{\boldsymbol{s}}

\newcommand{\bv}{\boldsymbol{v}}
\newcommand{\bx}{\boldsymbol{x}}

\newcommand{\bz}{\boldsymbol{z}}

\newcommand{\calD}{\mathcal{D}}

\newcommand{\calH}{\mathcal{H}}

\newcommand{\calN}{\mathcal{N}}

\usepackage[euler]{textgreek}
\usepackage[colorinlistoftodos]{todonotes}

\newcommand{\bmu}{\boldsymbol{\mu}}

\newcommand{\bSigma}{\boldsymbol{\Sigma}}

\newcommand{\vect}{\operatorname{vec}}

\newcommand{\blkdiag}{\operatorname{blkdiag}}

\usepackage{color}  
\def\cred{\textcolor{red}}

\def\cmag{\textcolor{magenta}}
\definecolor{darkgreen}{rgb}{0., 0.4, 0.}
\definecolor{amber}{rgb}{1.0, 0.49, 0.0}
\definecolor{orange}{rgb}{1.0, 0.4, 0.0}
\definecolor{darkorange}{rgb}{0.7, 0.32, 0.0}

\newcommand{\bPi}{\boldsymbol{\Pi}}

\renewcommand\cred{}

\renewcommand\cmag{}

\journal{ISPRS Journal of Photogrammetry and Remote Sensing}

\begin{document}

\begin{frontmatter}
\title{Online Fusion of Multi-resolution Multispectral Images with Weakly Supervised Temporal Dynamics}


\author[inst1]{Haoqing Li}
\author[inst1]{Bhavya Duvvuri}
\author[inst2]{Ricardo Borsoi}
\author[inst1]{Tales Imbiriba}
\author[inst1]{Edward Beighley}
\author[inst1]{Deniz Erdo{\u{g}}mu{\c{s}}}
\author[inst1]{Pau Closas}

\affiliation[inst1]{organization={Northeastern University},
            city={Boston},
            postcode={02215}, 
            state={MA},
            country={USA}}

\affiliation[inst2]{organization={University of Lorraine, CNRS, CRAN},
            city={Nancy},
            postcode={F-54000}, 
            country={France}}

\date{June 2022}

\begin{abstract}
Real-time satellite imaging has a central role in monitoring, detecting and estimating the intensity of key natural phenomena such as floods, earthquakes, etc. 
One important constraint of satellite imaging is the trade-off between spatial/spectral resolution and their revisiting time, a consequence of design and physical constraints imposed by satellite orbit among other technical limitations. 
In this paper, we focus on fusing multi-temporal, multi-spectral images where data acquired from different instruments with different spatial resolutions is used. 
We leverage the spatial relationship between images at multiple modalities to generate high-resolution image sequences at higher revisiting rates. To achieve this goal, we formulate the fusion method as a recursive state estimation problem and study its performance in filtering and smoothing contexts. Furthermore, a calibration strategy is proposed to estimate the time-varying temporal dynamics of the image sequence using only a small amount of historical image data. Differently from the training process in traditional machine learning algorithms, which usually require large datasets and computation times, the parameters of the temporal dynamical model are calibrated based on an analytical expression that uses only two of the images in the historical dataset.
A distributed version of the Bayesian filtering and smoothing strategies is also proposed to reduce its computational complexity.
\cmag{To evaluate the proposed methodology we consider a water mapping task where real data acquired by the Landsat and MODIS instruments are fused generating high spatial-temporal resolution image estimates. Our experiments show that the proposed methodology outperforms the competing methods in both estimation accuracy and water mapping tasks.}
\end{abstract} 



\end{frontmatter}


\section{Introduction}

High spatial resolution satellite image data is a fundamental tool for remote sensing applications such as the monitoring of land cover changes~\citep{lu2016land, zhu2014continuous}, deforestation~\citep{portillo2012forest,schultz2016performance} or water mapping~\citep{kim2017mapping, yoon2016estimating} and water quality~\citep{gholizadeh2016comprehensive}.
Moreover, to adequately deal with the variability of such events over time it is important to have short time spans between different image acquisitions of the same scene (i.e., a high temporal resolution, or low revisit times). 
However, fundamental limitations of multiband imaging instruments and large sensor-to-target distances impose a trade-off between spatial and temporal resolutions of satellite image sequences.

This means that instruments providing high spatial resolution have long revisit times, while the converse holds for instruments with short revisit times.
This can be illustrated, for instance, by considering Landsat~8 and MODIS instruments (with 30 and 250/500 meters spatial resolution, respectively). While MODIS is able to provide daily images at coarse resolution, Landsat-8 only revisits the same site once every 16 days~\citep{roy2014landsat}.

Considering these limitations, many works proposed multimodal image fusion techniques to generate high (spatial, spectral or temporal) resolution remote sensing images.
Multimodal image fusion aims to combine multiple observed images, each of which having high resolution in a given dimension -- spatial, temporal, or spectral -- to generate high resolution image sequences. 
\cred{
Several instances of image fusion have been considered, some works aim to directly supply classification maps from multiple satellite image and surface elevation data at each time instant~\citep{li2022domainKnowledgeFusionNetworkClassification}, integrating optical and radar data for time-series crop classification~\citep{yuan2023bridgingOpticalSARtimeSeriesCropClassification,garnot2022multiModalTimeSeriesCropMapping}, or fusing spatio-temporal optical and elevation data to obtain high-resolution land temperature maps~\citep{wu2022downscalingLandTemperature}.
}


\cred{In particular, classification or mapping tasks based on time-series remote sensing data is receiving increasing interest in the literature~\citep{sharma2018landCoverClassificationPatchRNNs,Fang2021,yuan2023bridgingOpticalSARtimeSeriesCropClassification,garnot2022multiModalTimeSeriesCropMapping}. Thus, to overcome the limitations of existing instruments,}
fusing images with different spectral and spatial resolutions has been extensively studied to generate images with high spatial and spectral resolutions, which are critical for accurately distinguishing different materials in a pixel~\citep{yokoya2017HS_MS_fusinoRev,Borsoi_2018_Fusion,loncan2015pansharpeningReview}. Recently, an increasing interest has been observed in applying multimodal image fusion to generate image sequences with high spatial and temporal resolutions~\citep{belgiu2019spatiotemporalFusionRev}, with particular interest dedicated to fusing data from multiple satellites to obtain daily images with high (e.g., 30~m) resolution~\citep{wang2018sentinel2_sentinel3_fusion}. This has already had an important impact in applications such as the generation of daily snow cover maps~\citep{rittger2021MEF_snowcover_30m} and the study of drought-induced tree mortality~\citep{yang2021forestMortality_HSR}.
Existing spatiotemporal image fusion methods are usually divided in weighted fusion, umixing-based, learning-based and Bayesian approaches~\citep{zhu2018spatiotemporalFusReview}. There also exist hybrid techniques, which leverage ideas from more than one family of approaches.


Weighted fusion methods assume that the temporal changes occurring between two time instants are consistent between the high and low spatial resolution images for low resolution pixels which are composed of only a single material~\citep{gao2015fusingLandsatMODISreview}. However, coarse resolution pixels are often mixtures of different materials. The predicted high resolution pixels are then computed as a weighted linear combination of the previous high resolution pixels and of the changes occurring at low resolution pixels in a given neighborhood~\citep{gao2006STARFM,zhu2010fusion_ESTARFM}. Different works have designed various weighting functions, which aim to select neighboring pixels that are homogeneous and spatially/spectrally similar to the pixel whose change is being predicted~\citep{gao2006STARFM,zhu2018spatiotemporalFusReview,zhang2018spatiotemporalFusionWeightedMultiscale}.
Other works have extended such framework account for sudden changes~\citep{hilker2009STAARCH_fusion} or to use different weighting functions~\citep{zhu2010fusion_ESTARFM}.


\begin{figure*}[htb]
    \centering
    \includegraphics[width=0.8\textwidth]{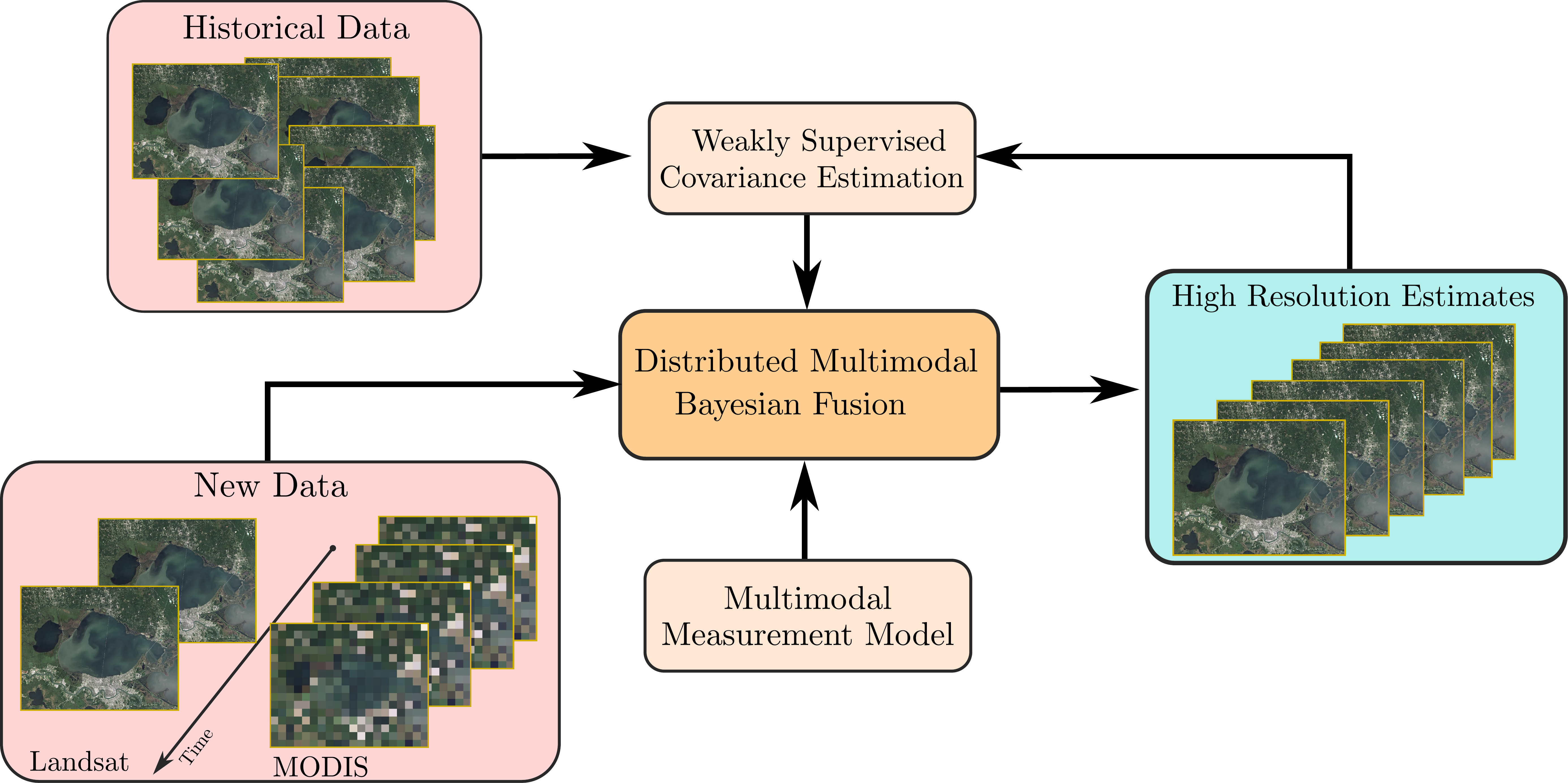}
    \caption{Overview of the proposed method. Multimodal (e.g., Landsat and MODIS over time) images time series  are fused by the Distributed Multimodal Bayesian Fusion algorithm resulting in a high spatial-temporal resolution estimated sequence. Covariance estimates for the dynamical model are estimated through a  weakly supervised strategy based on local high-resolution historical data. We highlight that the Bayesian fusion  methodology employed here is agnostic to the multimodal measurement model making the strategy easily generalizable to different data scenarios. }
    \label{fig:overview}
\end{figure*}   

Unmixing-based methods make use of the linear mixing model (LMM), which assumes that each pixel in the low resolution image can be represented as a convex combination of the reflectance of a small number of pure spectral signatures, called endmembers~\citep{keshava2002unmixingReview,borsoi2018superpixels1_sparseU}.
The LMM has been used for multimodal image fusion by assuming the proportions of each material in a low resolution pixel to be stable/constant over time~\citep{zurita2008unmixingFusion1,amoros2013multitemporalFusionUnmixing,wu2012unmixingFusion2}. This way, spectral unmixing~\citep{keshava2002unmixingReview} is used to estimate the endmembers at different time instants from low resolution images, while using different strategies to mitigate the spectral variability of a single material~\citep{zurita2008unmixingFusion1,borsoi2020variabilityReview,Borsoi_multiscaleVar_2018}. However, abrupt abundance variations (originating from, e.g., land cover changes) are commonly found in multitemporal image streams~\citep{li2020sfsdaf_enhanced_spatiotemporal_fusion,liu2019reviewCD,borsoi2021MT_MESMA,erturk2015sparseSU_CD}, which may negatively impact the performance of such methods and can be particularly challenging to address when occurring jointly with finer endmember variations~\citep{li2020sfsdaf_enhanced_spatiotemporal_fusion}. Thus, special care is required when fusing images which are temporally distant from one another~\citep{wang2020virtualImagePairSpatiotemporalFusion}, motivating the development of strategies using, e.g., spatially adaptive quantification of the reliability of the input images to guide unmixing based image fusion strategies~\citep{shi2022reliableSpatiotemporalFusion}.


Learning-based approaches leverage training data and different machine learning algorithms in order to perform image fusion. Those approaches are varied, ranging from approaches such as dictionary learning~\citep{huang2012spatiotemporalFusionDictLearning}, which are based on a sparse representation of image pixels and have a strong connection to the LMM, to convolutional neural networks~\citep{song2018spatiotemporalFusionCNNs}, which are flexible function approximations which are typically used to learn a mapping from the low-resolution to high resolution data.


Bayesian methods are flexible alternatives to the previous approaches that take into account the uncertainty present both in the imaging model and in the estimated images. 
The Bayesian framework is based on the definition of probabilistic models to describe the relationship between images of different spatial, spectral and temporal resolutions acquired by different instruments. This allows image fusion to be formulated as a maximum a posteriori estimation problem~\citep{shen2016integratedSpatioTemporalFusion}.
Although Bayesian methods usually consider Gaussian distributions for mathematical tractability, different variations have been proposed depending on how the image acquisition process is modelled and on how the mean and covariance matrices are estimated. This included assuming them diagonal~\citep{huang2013spatiotemporalFusionBayes}, estimating image covariance matrices based on an initial estimate of the high resolution image~\citep{shen2016integratedSpatioTemporalFusion}, or based on the low resolution image pixels~\citep{xue2017bayesianFusionPixelCovariances}.

A recent work considered a Kalman filter-based approach to estimate a high resolution image sequence based on mixed resolution observations from the Landsat and MODIS instruments~\citep{zhou2020kalmanFusionLandsatMODIS}. However, to define the model for the Kalman filter, two Landsat+MODIS image pairs at times $t_0$ and $t_N$ are considered, as well as a time series of MODIS images at instants $t_k\in[t_0,t_N]$, making it unsuitable for online operation. Moreover, changes between each pair of images were assumed to be constant/uniform over predefined groups of high resolution image pixels, which can be restrictive (due to the large resolution difference between the measured images, the groups must contain many pixels in order to make the model well-posed). It also does not benefit from auxiliary information that could aid the estimation of the high resolution images.
Another work used the Kalman filter to estimate normalized difference vegetation indices (NDVI) time series images from Landsat and MODIS observations, using an affine model for the dynamics of the states whose coefficients are selected based on the seasonality, and another affine model to relate the NVDI estimate obtained from MODIS and Landsat measurements~\citep{sedano2014kalmanFUsionNDVI}. The Kalman filter was also recently applied to estimate land surface temperature by fusing thermal infrared and microwave data~\citep{xu2021landTemperatureFusionKalman}.



In this paper, we propose a weakly supervised Kalman filter and smoother framework for spatio-temporal fusion of multispectral images. The proposed framework relies on explicit modeling assumptions about the image acquisition and temporal evolution processes, under which the proposed solution is statistically optimal. The Kalman filter-based methods can operate in a fully online setting, where high-resolution images are only available as past data.
We also develop a smoother-based method to optimally exploit information contained in future high-resolution observed images when processing images in a time window. 
However, the quality of the reconstruction of Kalman filter and smoother strategies depend directly on the quality of the dynamical image evolution model. Thus, to overcome this limitation, a weakly supervised strategy is proposed to learn the temporal dynamics of the high-resolution images from a small amount of past data. More precisely, instead of considering the changes to be constant over areas comprising large amounts of image pixels, we propose an analytical calibration strategy to estimate a more informative time-varying dynamical image model by leveraging historical data. This allows for a better localization of changes in the high resolution image even in intervals where only coarse resolution observations (e.g., MODIS) are available. Moreover, to mitigate the high computational complexity of the Kalman filter and smoother, we propose a distributed implementation by exploiting different independence assumptions about the high-resolution state space, allowing the proposed methods to be applied to large datasets and geographical areas.
Figure~\ref{fig:overview} depicts the proposed methodology where high-resolution (spatially and temporally) estimates are generated by fusing different data modalities. 
We illustrate the application of the proposed framework by fusing images from the Landsat and MODIS instruments. Experimental results indicate that the proposed method can lead to considerable improvements compared to using a non-informative dynamical model and to widely used image fusion algorithms, both in image reconstruction and in downstream water classification and hydrograph estimation tasks.
\cred{A software package containing an implementation of the proposed method and the image dataset is available at \url{https://github.com/HaoqingLi/Multi-resolution-Multispectral-image-fusion-based-weakly-supervised-constrained-Kalman-filter}.}

This paper is organized as follows. In Section~\ref{sec:model}, we present the paper notation and the proposed imaging model. Section~\ref{sec:filter} presents the Kalman filter and smoother approaches for multimodal image fusion. Section~\ref{sec:experiments} contains simulation experiments that illustrate the performance of the proposed method. Finally, Section~\ref{sec:conclusions} concludes the paper.














\section{Dynamical Imaging Model}
\label{sec:model}

\subsection{Definitions and notation}

Let us denote the the $\ell$-th band of the $k$-th acquired image reflectances from modality $m\in\Omega$ by $\by_{k,\ell}^{m}\in{\amsmathbb{R}}^{N_{m,\ell}}$, with $N_{m,\ell}$ pixels for each of the bands $\ell=1,\ldots,L_m$, and $\Omega$ denoting the set of image modalities. 
As a practical example, we consider $\Omega=\{\mathsf{L},\mathsf{M}\}$ to contain the Landsat-8, and MODIS image modalities, without loss of generality. 
We also denote by $\Omega_H$ the highest resolution image modality, e.g., $\Omega_H=\{\mathsf{L}\}$.
%
%
We denote the corresponding high resolution latent reflectances by $\bS_k\in{\amsmathbb{R}}^{N_H\times L_H}$, with $N_H$ pixels and $L_H$ bands, with $L_H\geq L_m$ and $N_H\geq N_{m,\ell}$, $\forall \ell,m$. Subindex $k\in{\amsmathbb{N}}_*$ denotes the acquisition time index. We also denote by $\operatorname{vec}(\cdot)$, $\operatorname{col}\{\cdot\}$, $\operatorname{diag}\{\cdot\}$ and by $\operatorname{blkdiag}\{\cdot\}$ the vectorization, vector stacking, diagonal and block diagonal matrix operators, respectively. The notation $\bx_{a:b}$ for $a,b\in{\amsmathbb{N}}_*$ represents the set $\{\bx_a,\bx_{a+1},\ldots,\bx_b\}$. We use ${\calN}(\bmu,\bSigma)$ to denote a Gaussian distribution with mean $\bmu$ and covariance matrix $\bSigma$.

\subsection{Measurement model}
To formulate our measurement model we assume that the acquired image at time index $k$, for any imaging modality, is a spatially degraded and spectrally transformed version of the high resolution latent reflectance image $\bS_k$.
Following this assumption our measurement model for the $m$-th modality becomes:
\begin{align}
    \by_{k,\ell}^{m} = {\calH}_{\ell}^{m}(\bS_k)\bc_{\ell}^{m} + \br_{k,\ell}^{m} \,, \qquad \ell=1,\ldots,L_m \,,
    \label{eq:obs_mdl_1}
\end{align}
where $\bc_{\ell}^{m}\in{\amsmathbb{R}}^{L_H}$ denotes a spectral transformation vector, mapping all bands in $\bS_k$ to the $\ell$-th measured band at modality $m$; ${\calH}_{\ell}^{m}$ is a linear operator representing the band-wise spatial degradation, modeling blurring and downsampling effects of each high resolution band, and $\br_{k,\ell}^{m}$ represents the measurement noise. Note that, while we consider the spatial resolution of the high resolution bands in $\bS_k$ to be the same, different bands from the same modality can have different resolutions. We also assume the measurement noise to be Gaussian and uncorrelated among bands, that is, $\br_{k,\ell}^{m}\sim{\calN}(\cb{0},\bR_{\ell}^m)$ with time-invariant covariance matrix given by $\bR_{\ell}^m\in{\amsmathbb{R}}^{N_{m,\ell}\times N_{m,\ell}}$, and $\operatorname{cov}(\br_{k,j}^{m},\br_{k,\ell}^{m})=\cb{0}$ for all $j\neq\ell$.
%

Note that satellite images may be corrupted by several effects, including dead pixels in the sensor, incorrect atmospheric compensation, and the presence of heavy cloud cover. Such pixels cannot be reliably used in the image fusion process as they may degrade the performance of the method. Directly addressing these effects using a statistical model would require the choice of a non-Gaussian distribution for the noise vector $\br_{k,\ell}^{m}$, which could make the computational complexity of the fusion procedure prohibitive.
Thus, we consider a matrix $\bD_k^{m}\in{\amsmathbb{R}}^{\widetilde{N}_m\times N_m}$, which eliminates outlier pixels from the image, leading to the following transformed measurement model:
\begin{align}
    \widetilde{\by}_{k,\ell}^{m} =  \bD_k^{m}{\calH}_{\ell}^{m}(\bS_k)\bc_{\ell}^{m} + \widetilde{\br}_{k,\ell}^{m} \,,
    \label{eq:obs_mdl_2}
\end{align}
where $\widetilde{\by}_{k,\ell}^{m}=\bD_k^{m}\by_{k,\ell}^{m}$ and $\widetilde{\br}_{k,\ell}^{m}=\bD_k^{m}\br_{k,\ell}^{m}$ denotes the measured image band and the measurement noise in which the outlier values have been removed.

Using~\eqref{eq:obs_mdl_2} and the properties of the vectorization operator, we can write this model equivalently as
\begin{align}
    \widetilde{\by}_{k,\ell}^{m}
    &= \big[(\bc_{\ell}^{m})^\top \otimes \bD_k^{m} \big]  \vect\big({\calH}_{\ell}^{m}(\bS_k)\big) + \widetilde{\br}_{k,\ell}^{m}
    \nonumber \\
    &= \big[(\bc_{\ell}^{m})^\top \otimes \bD_k^{m} \big]  \bH_{\ell}^{m}\bs_k + \widetilde{\br}_{k,\ell}^{m}
\end{align}
where $\otimes$ denotes the Kronecker product.
%
The variable $\bs_k\in\amsmathbb{R}^{L_H N_H}$ denotes a vector-ordering of the high-resolution image $\bS_k$ which is obtained by grouping all pixels such that the bands of a single HR pixel are adjacent to each other, and the pixels that are contained within a single ``lowest-resolution'' pixel are also adjacent to each other, that is:
\begin{align}
    \bs_k = \begin{bmatrix}
    \begin{bmatrix}
    s_{k,1,\iota(1,1)} \\ \vdots \\ s_{k,L_H,\iota(1,1)} \\ s_{k,1,\iota(2,1)} \\ \vdots \\ s_{k,L_H,\iota(d,1)} 
    \end{bmatrix}^\top,
    \ldots,
    \begin{bmatrix}
    s_{k,1,\iota(1,N_{m',\ell'})} \\ \vdots \\ s_{k,L_H,\iota(1,N_{m',\ell'})} \\ s_{k,1,\iota(2,N_{m',\ell'})} \\ \vdots \\ s_{k,L_H,\iota(d,N_{m',\ell'})} 
    \end{bmatrix}^\top
    \end{bmatrix}^\top \,,
\end{align}
where $s_{k,i,j}$ is the $(i,j)$-th position of $\bS_k$, $m'$ and $\ell'$ are the modality and spectral band with the lowest spatial resolution (i.e., for which $N_{m,\ell}$ is smallest), $d=N_H/N_{m',\ell'}$ is the number of HR pixels inside each low resolution pixel of band $\ell'$ and modality $m'$, and $\iota:\amsmathbb{N}_*\times\amsmathbb{N}_*\to\amsmathbb{N}_*$ is a function such that $\iota(i,j)$ returns the index (in $\bS_k$) of the of the $i$-th HR pixel contained inside the $j$-th low resolution pixel (where $i\in\{1,\ldots,d\}$) for modality $m'$ and band $\ell'$. $\bH_{\ell}^{m}$ is a matrix form representation of the operator ${\calH}_{\ell}^{m}$, such that $\vect({\calH}_{\ell}^{m}(\bS_k))=\bH_{\ell}^{m}\bs_k$.


We can now represent all bands from each modality in the form of a single vector $\widetilde{\by}_{k}^{m}\in{\amsmathbb{R}}^{\widetilde{N}_mL_m}$ as
\begin{align}
    \widetilde{\by}_{k}^{m} = 
    \underbrace{\begin{pmatrix}
    \big[(\bc_{1}^{m})^\top \otimes \bD_k^{m} \big] \bH_{1}^{m} \\
    \vdots\\
    \big[(\bc_{L_m}^{m})^\top \otimes \bD_k^{m} \big] \bH_{L_m}^{m}
    \end{pmatrix}}_{\widetilde{\bH}_k^{m}} \bs_k + \widetilde{\br}_{k}^{m} \,,
    \label{eq:meas_mdl_4}
\end{align}
where $\widetilde{\br}_{k}^{m}\sim{\calN}(\cb{0},\widetilde{\bR}_k^m)$, and
\begin{align}
    \widetilde{\by}_{k}^{m} &= \operatorname{col}\big\{\widetilde{\by}_{k,1}^{m},\ldots,\widetilde{\by}_{k,L_m}^{m}\big\} \,,
    \\
    \widetilde{\br}_{k}^{m} &= \operatorname{col}\big\{\widetilde{\br}_{k,1}^{m},\ldots,\widetilde{\br}_{k,L_m}^{m}\big\} \,,
    \\
    \widetilde{\bR}_k^m &= \operatorname{blkdiag}\big\{\bD_k^{m}\bR_{1}^m(\bD_k^{m})^\top, \ldots, \bD_k^{m}\bR_{L_m}^m(\bD_k^{m})^\top\big\}
\end{align}
Note that at most time instants $k$, one or more of the modalities $m\in\Omega$ is not observed. In this case, we set the matrix $\bD_k^{m}$ as an empty (zero-dimensional) matrix, which simplifies the problem and avoids introducing additional notation.


\subsection{Dynamical evolution model}
Defining reasonable dynamical models for image fusion requires detailed knowledge regarding the scene evolution over time, which is often unattainable. 
In this contribution, we aim at a complete data driven strategy assuming very little knowledge regarding the scene evolution except for past data coming from the imaging modalities being used. To match such lack of prior knowledge we consider a simple random-walk process to model the latent state dynamics as:
\begin{align}
    \bs_{k+1} = \bF_k\bs_k + \bq_k \,,
    \label{eq:state_evol_mdl_1}
\end{align}
where $\bF_k\in{\amsmathbb{R}}^{L_H N_H \times L_H N_H}$ is the state transition matrix, which is assumed to satisfy $\|\bF_k\|_2\leq1$, and $\bq_k\sim{\calN}(\cb{0},\bQ_k)$ with $\bQ_k\in{\amsmathbb{R}}^{L_H N_H\times L_H N_H}$ being the state process noise covariance matrix.
Note that the above model plays a crucial role in the estimation results, as it describes both the distribution of the changes occurring in the image at time $k$, as well as the marginal distribution of the states. 
This means that more sophisticated dynamics can be introduced in the problem through the appropriate design of the process noise covariance matrix $\bQ_k$.
Although expectation maximization (EM) can be used to estimate $\bQ_k$ in time invariant models~\citep{borsoi2020multitemporalUKalmanEM}, the problem becomes extremely ill-posed in the time-varying setting. 
Another issue relates to the computational complexity of EM-based strategies requiring the solution of the Kalman filter and smoother systems multiple times, becoming unfeasible when dealing with large images. 
%
For these reasons, we propose an alternative route to estimate $\bQ_k$.

\subsection{A weakly supervised approach for estimating $\bQ_k$}\label{sec:Qest}

We consider $\bQ_k(\cp{D}_k)$ as a function of the set  $\cp{D}_k=\{\tilde{\by}_\ell^{m\in\Omega_H}\}_{\ell<k}$ of past high resolution images. The set $\cp{D}_k$ represents historical data and images currently being fused up the the time step $k$.
Although many strategies could be leveraged to find suitable past time windows to account for more relevant covariance estimation and consider full covariance matrices, in this preliminary work we choose a simple route to validate this type of approach. 
For this, let ${\by}^{m\in\Omega_H}_{k-\tau}$ be the the most recently observed high resolution image\footnote{That is, $\tau\in{\amsmathbb{Z}_+}$ is the smallest integer such that a high resolution image was observed at time instant $k-\tau$.}. We compute $\bQ_k$ by finding in our historical data the most similar image to ${\by}^{m\in\Omega_H}_{k-\tau}$ and then computing the pixelwise variance across the following $n\in\amsmathbb{N}_*$ images in our historical data. That is, we compute $\bQ_k$ executing the following three steps for every time step $k$:
\begin{enumerate}
    \item Identify the most similar state over $\cp{D}_k$, that is, the image that is most similar, according to a metric $\cal{L}$
    \begin{equation}
        \ell^* = \mathop{\arg\min}_{\ell \in {\cp{I}}_{\cp{D}_k}} \,\, {\cal{L}}\big({\by}^{m\in\Omega_H}_{k-\tau}, [{\cal{D}}_k]_\ell\big) \,,
    \end{equation}
    with $[{\cal{D}}_k]_\ell$ being the $\ell$-th image in the historical set ${\cal{D}}_k$, and ${\cp{I}}_{\cp{D}_k}\subseteq{\amsmathbb{Z}}$ is the set containing the time index of each image in ${\cp{D}_k}$.
    
    \item select a time window $[{{\calD}_k}]_{\ell^*:\ell^*+n}$.
    
    \item compute the diagonal process noise covariance matrix, i.e.,  $\bQ_k=\operatorname{diag}\{q_{k,1}^2, \ldots, q^2_{k,L_HN_H}\}$, as
    \begin{equation}
    q^2_{k,j} = \max\bigg(\frac{\operatorname{var}\big([{\cal{D}}_k]_{\ell^*:\ell^* + n}^{(j)}\big)}{\Delta_{{\cal{D}}_k}^{\ell^*}}, \varepsilon^2\bigg) \times \Delta_k\,,
\end{equation}
\end{enumerate}
where $[{\cal{D}}_k]_{\ell^*:\ell^*+n}^{(j)} = [\tilde{y}_{\ell^*,j}^{m\in\Omega_H},\ldots, \tilde{y}_{\ell^*+n,j}^{m\in\Omega_H}]$, $\varepsilon>0$ is a small scalar allowing for changes on the scene that were unseen on the historical data window $[{\cal{D}}_k]_{\ell^*:\ell^*+n}$, $\Delta_k$ is the time interval (in days) between ${\by}^{m\in\Omega_H}_{k}$ and ${\by}^{m\in\Omega_H}_{k+1}$, and $\Delta_{{\cal{D}}_k}^{\ell^*}$ is the time interval (in days) between $[{\cal{D}}_k]_{\ell^*}$ and $[{\cal{D}}_k]_{\ell^*+n}$.
As similarity metric we used the cosine similarity $\cal{L}(\by, \bz) = \cos(\by,\bz)$.

\section{Multimodal image fusion using a weakly supervised constrained Kalman filter}
\label{sec:filter}

Considering models \eqref{eq:meas_mdl_4} and \eqref{eq:state_evol_mdl_1}, the online multimodal image fusion problem can be formulated as the problem of computing the posterior distribution of the high resolution image given all previous measurements available, i.e.,
\begin{align}
    p\big(\bs_k\big|\{\widetilde{\by}_{1:k}^{m}\}_{m\in\Omega}\big) = {\calN}\big(\bs_{k|k},\bP_{k|k}\big)\,.
    \label{eq:posterior_1}
\end{align}
%
Due to the choice of a linear Gaussian model, this distribution is also Gaussian. Moreover, its mean vector $\bs_{k|k}$ and covariance matrix $\bP_{k|k}$ can be computed recursively using the standard Kalman filter with a prediction and update steps~\citep{sarkka2013bayesianBook}.


More precisely, the prediction step of the Kalman filter computes the first and second order moments of $p\big(\bs_k\big|\{\widetilde{\by}_{1:k-1}^{m}\}_{m\in\Omega}\big)$ as:
\begin{align}
    \bs_{k|k-1} &= \bF_{k-1}\bs_{k-1|k-1}
    \label{eq:kalman_pred1}
    \\
    \bP_{k|k-1} &= \bF_{k-1}\bP_{k-1|k-1}\bF_{k-1}^\top + \bQ_{k-1}
    \label{eq:kalman_pred2}
\end{align}
The update step computes then computes of~\eqref{eq:posterior_1}. Note that the update step can be simplified and implemented separately for each data modality by using the Markov property of the model and the independence between noise vectors of different modelities:
\begin{align}
    p\big(\bs_k\big|&\{\widetilde{\by}_{1:k}^{m}\}_{m\in\Omega}\big)
    \nonumber \\
    & \propto
    p\big(\{\widetilde{\by}_k^{m}\}_{m\in\Omega}\big|\bs_k\big) 
    \nonumber 
    p\big(\bs_k\big|\{\widetilde{\by}_{1:k-1}^{m}\}_{m\in\Omega}\big)
    \\
    & = p\big(\bs_k\big|\{\widetilde{\by}_{1:k-1}^{u}\}_{u\in\Omega}\big) \prod_{m\in\Omega} p\big(\widetilde{\by}_k^{m}\big|\bs_k\big) \,.
    \label{eq:multi_upd_1}
\end{align}
By computing the first product in the right hand side as:
\begin{align}
    & p\big(\bs_k\big|\{\widetilde{\by}_{1:k-1}^{u}\}_{u\in\Omega}\big) p\big(\widetilde{\by}_k^{m}\big|\bs_k\big) 
    \nonumber\\
    & \hspace{7ex} \propto p\big(\bs_k\big|\{\widetilde{\by}_{1:k-1}^{u}\}_{u\in\Omega},\widetilde{\by}_k^{m}\big) \,,
    \label{eq:multi_upd_2}
\end{align}
which is an update step of the Kalman filter with image modality $m$ to yield a new posterior in the r.h.s. of~\eqref{eq:multi_upd_2}. This can be computed as:
\begin{align}
    \bv_k^m &= \widetilde{\by}_{k}^{m} - \widetilde{\bH}_k^{m} \bs_{k|k-1}
    \label{eq:kalman_up1}\\
    \bT_k^m &= \widetilde{\bH}_k^{m}\bP_{k|k-1}\big(\widetilde{\bH}_k^{m}\big)^\top + \widetilde{\bR}_k^m
    \label{eq:kalman_up2}\\
    \bK_k^m &=  \bP_{k|k-1} \big(\widetilde{\bH}_k^{m}\big)^\top \big(\bT_k^m\big)^{-1} 
    \label{eq:kalman_up3}\\
    \bs_{k|k} &= \bs_{k|k-1} + \bK_k^m \bv_k^m
    \label{eq:kalman_up4}\\
    \bP_{k|k} &=  \bP_{k|k-1} - \bK_k^m\bT_k^m \big(\bK_k^m\big)^\top
    \label{eq:kalman_up5}
\end{align}
for $m\in\Omega$. By proceeding with the computation of the product in the r.h.s. of~\eqref{eq:multi_upd_1} recursively, the Kalman update can then be performed separately for each of the modalities observed at time instant~$k$. Note that after the first modality is processed, the update equations above are used again for the subsequent modalities by setting $\bs_{k+1|k}$ and $\bP_{k+1|k}$ as equal to the posterior estimates from the previously processed modality.

\subsection{The Linear Smoother}

Given a window of $K$ image samples, the Bayesian smoothing problem consists of computing the posterior distribution of the high resolution image given all available measurements available, i.e.,
\begin{align}
    p\big(\bs_k\big|\{\widetilde{\by}_{1:K}^{m}\}_{m\in\Omega}\big) = {\calN}\big(\bs_{k|K},\bP_{k|K}\big) \,,
    \label{eq:posterior_2}
\end{align}
which is also a Gaussian.
Just like in the filtering problem, the linear and Gaussian model allows this solution to be computed efficiently using the Rauch-Tung-Striebel (RTS) smoothing equations~\citep{sarkka2013bayesianBook}, which consist of a forward pass of the Kalman filter (as described before), followed by a backwards recursion that updates the previously computed mean and covariances matrices of the state with information from future time instants.

We note that the smoothing can also be performed efficiently for the case when multiple image modalities are available. Let us consider the Bayesian smoothing equations as defined in~\citep{kitagawa1987smoother,sarkka2013bayesianBook}, which is performed in two steps. Starting from the Kalman state estimate at time $K$, given by $p\big(\bs_{K}\big|\{\widetilde{\by}_{1:K}^{m}\}_{m\in\Omega}\big)$, the smoothing distribution is computed recursively for $k=k-1,\ldots,1$, according to the following relation:
\begin{align}
    p\big(\bs_{k}\big|\{&\widetilde{\by}_{1:K}^{m}\}_{m\in\Omega}\big) = p\big(\bs_k\big|\{\widetilde{\by}_{1:k}^{m}\}_{m\in\Omega}\big)
    \nonumber\\
    & \times \int \frac{p(\bs_{k+1}|\bs_k)p\big(\bs_{k+1}\big|\{\widetilde{\by}_{1:K}^{m}\}_{m\in\Omega}\big)}{p\big(\bs_{k+1}\big|\{\widetilde{\by}_{1:k}^{m}\}_{m\in\Omega}\big)} d\bs_{k+1} \,,
\end{align}
where $p\big(\bs_k\big|\{\widetilde{\by}_{1:k}^{m}\}_{m\in\Omega}\big)={\calN}(\bs_{k|k},\bP_{k|k})$ is the Kalman estimate of the state PDF at time $k$, $p(\bs_{k+1}|\bs_k))$ is the state transition PDF, computed according to~\eqref{eq:state_evol_mdl_1}, $p\big(\bs_{k+1}\big|\{\widetilde{\by}_{1:K}^{m}\}_{m\in\Omega}\big)={\calN}(\bs_{k+1|K},\bP_{k+1|K})$ is the smoothing distribution obtained at the previous iteration, and $p\big(\bs_{k+1}\big|\{\widetilde{\by}_{1:k}^{m}\}_{m\in\Omega}\big)$ is the predictive state distribution, which is computed exactly as in the prediction step of the Kalman filter.

In the linear and Gaussian case this translates into the following closed form solution~\citep{sarkka2013bayesianBook}, with 
\begin{align}
    \bs_{k+1|k} &= \bF_k\bs_{k|k} 
    \label{eq:smoother_pred1}
    \\
    \bP_{k+1|k} &= \bF_k\bP_{k|k}\bF_k^\top + \bQ_k
    \label{eq:smoother_pred2}
\end{align}
being used to compute the predictive state distribution, and
\begin{align}
    \bG_k &= \bP_{k|k}\bF_k^\top \bP_{k+1|k}^{-1}
    \label{eq:smooth_up1}
    \\
    \bs_{k|K} &= \bs_{k|k} + \bG_k(\bs_{k+1|K} - \bs_{k+1|k})
    \label{eq:smooth_up2}
    \\
    \bP_{k|K} &= \bP_{k} + \bG_k(\bP_{k+1|K} - \bP_{k+1|k}) \bG_k^\top
    \label{eq:smooth_up3}
\end{align}
to update the covariances. It should be noted that the mean and covariance $\bs_{k|k}$ and $\bP_{k|k}$ used in the Smoothing equations are the final result obtained from the Kalman update after processing all image modalities that were available at instant~$k$.

Thus, while in the Kalman filtering the update equations must be computed sequentially at each time step w.r.t. the different image modalities, smoothing only needs only the final state estimates at each instant, no matter how many modalities are present.

\subsection{Constraining the estimates}

Although the Kalman filter provides closed-form solutions to the estimation of the high-resolution image sequence, it relies on a Gaussian assumption on the states and observations which does not correspond to the physics of the problem. In fact, represented in reflectance values, each pixel and band of a high-resolution images $\bs_k$ is actually constrained to an interval $s_{k,i,j}\in[0,s_{\max}]$, where $s_{\max}$ is the maximum reflectance values of the scene. Since this information can potentially improve the accuracy of the estimated states, we propose to incorporate this information by considering the linearly constrained Kalman filter~\citep{simon2010constrained_kalman}, in which the final constrained state $\bs_{k|k}^+$ is obtained as the solution to a constrained optimization problem:
\begin{align}
\begin{split} \label{eq:constrained_KF_upd}
    \bs_{k|k}^+ = & \arg\min_{\bs} \,\,\, \big(\bs-\bs_{k|k}\big)^\top\bP_{k|k}^{-1}\big(\bs-\bs_{k|k}\big)
    \\
    & \text{subject to } \, \bs\in[0,s_{\max}]^{N_HL_H}
\end{split} \,.
\end{align}
Problem~\eqref{eq:constrained_KF_upd} consists in a constrained quadratic program, which can be costly to solve due to the high dimensionality of the variables. Thus, we propose a simple solution consisting of truncating the result of the traditional Kalman update:
\begin{align}
    \bs_{k|k}^+ = \max\big(\min\big(\bs_{k|k},s_{\max}\big),0\big) \,,
    \label{eq:constrained_kf_simple}
\end{align}
where functions $\max(\cdot,\cdot)$ and $\min(\cdot,\cdot)$ compute the elementwise maximum and minimum value between a vector and a scalar. Note that this truncation provides the exact solution when $\bP_{k|k}$ is diagonal. The same truncation strategy was also applied to the results of the linear smoother $\bs_{k|K}$.
We generally observed that this gave good results in practice. $s_{\max}$ can be estimated as the maximum value of the observed images in a time window, or from the historical data.

\section{A distributed implementation}\label{sec:complex}

A problem with the Kalman filter is the need to compute and store the state covariance matrix, $\bP_{k|k}$. This incurs in storage and operations asymptotic complexity in the order of $\mathcal{O}(N_H^2 L_H^2)$ and $\mathcal{O}(N_H^3 L_H^3)$, respectively. This can make the method intractable for images with a large number of pixels. Thus, to reduce the complexity of the filter and of the smoother, we consider splitting the pixels in the estimated state $\bs_k$ into multiple groups which are assumed to be statistically independent~\citep{closas2012multiple,vila2016uncertaintyMultipleKalman,vila2017multiple}.
To this end, we divide the state space into $G$ groups as:
\begin{align}
    \bs_k = \vect\big([\bs_k^{(1)},\ldots,\bs_k^{(G)}]\big) \,,
    \label{eq:dist_kf_groups}
\end{align}
where the variables within each block $\bs_k^{(g)}$ are correlated, but different blocks $\bs_k^{(g_1)}$ and $\bs_k^{(g_2)}$ are assumed to be independent for $g_1\neq g_2$. This leads to the following approximation for the predictive and posterior covariance matrices $\bP_{k|k-1}$ and $\bP_{k|k}$ as block diagonal matrices:
\begin{align}
    \bP_{k|k-1} &= \blkdiag\Big\{\bP_{k|k-1}^{(1)},\ldots,\bP_{k|k-1}^{(G)} \Big\}
    \label{eq:dist_kf_groups_Pkold}
    \\
    \bP_{k|k} &= \blkdiag\Big\{\bP_{k|k}^{(1)},\ldots,\bP_{k|k}^{(G)} \Big\}
    \label{eq:dist_kf_groups_Pk}
\end{align}

We consider different splitting possibilities, with different trade-offs between approximation accuracy with respect to the full-state-covariance Kalman filter and complexity:
\begin{itemize}
    \item[$i)$] A fully diagonal model (with $G=N_H L_H$ blocks).
    \item[$ii)$] A block diagonal model where each block consists of all bands of one single high-resolution pixel (with $G=N_H$ blocks).
    \item[$iii)$] A block diagonal model, with blocks corresponding to the high-resolution pixels which reside inside a single MODIS pixel (with $G=N_HL_H/N_{\text{MODIS}}$ blocks).
\end{itemize}

Following~\citep{vila2016uncertaintyMultipleKalman}, the Kalman equations for the prediction step~\eqref{eq:kalman_pred1}--\eqref{eq:kalman_pred2} can be written for each block as:
\begin{align}
    \bs_{k+1|k}^{(g)} &= \big[\bF_k\big]_{(g),:} \bs_k
    \label{eq:kalman_pred1_dist}
    \\
    \bP_{k+1|k}^{(g)} &= \big[\bF_k\big]_{(g),:} \bP_{k} \big(\big[\bF_k\big]_{(g),:}\big)^\top + \bQ_k^{(g)}
    \label{eq:kalman_pred2_dist}
\end{align}
where $\big[\bF_k\big]_{(g),:}$ means the matrix formed by taking from $\bF_k$ the rows which correspond to the indices in the group of states $g$, and all columns. Matrices $\bQ_k^{(g)}$ are defined as:
\begin{align}
    \bQ_k = \blkdiag\big\{ \bQ_k^{(1)},\ldots,\bQ_k^{(G)} \big\} \,.
\end{align}
Similarly, the Kalman update equations~\eqref{eq:kalman_up1}--\eqref{eq:kalman_up5} are performed separately for each block of variables, and are given by:
\begin{align}
    %
    \bs_k^{(g)} &= \bs_{k|k-1}^{(g)} + \bK_k^{(g)} \bv_k^m
    \label{eq:kalman_dist_up1}
    \\
    \bP_{k}^{(g)} &= \bP_{k|k-1}^{(g)} - \bK_k^{(g)} \bT_k^m  \big(\bK_k^{(g)}\big)^\top
    \label{eq:kalman_dist_up2}
\end{align}
with:
\begin{align}
    \bK_k^{(g)} &= \bSigma_{xy,k|k-1}^{(g)} \big(\bT_k^m\big)^{-1}
    \label{eq:kalman_dist_up3}
    \\
    \bv_k^m &= \widetilde{\by}_{k}^{m} - \widetilde{\bH}_k^{m} \bs_{k|k-1}
    \label{eq:kalman_dist_up4}
    \\
    \bT_k^m &= \widetilde{\bH}_k^{m} \bP_{k|k-1} \big(\widetilde{\bH}_k^{m}\big)^\top + \widetilde{\bR}_k^m
    \label{eq:kalman_dist_up5}
    \\
    \bSigma_{xy,k|k-1}^{(g)} &= \big[\bP_{k|k-1}\big(\widetilde{\bH}_k^{m}\big)^\top\big]_{(g),:} 
    \nonumber \\
    & = \big[\bP_{k|k-1}\big]_{(g),:} \big(\widetilde{\bH}_k^{m}\big)^\top
    \label{eq:kalman_dist_up6}
\end{align}
where $\big[\bP_{k|k-1}\big]_{(g),:}$ means the matrix formed by taking from $\bP_{k|k-1}$ the rows which correspond to the indices in the group of states $g$, and all columns. Note that the block diagonal structure of $\bP_{k|k-1}$ and $\bP_{k|k}$ can be explored to perform the above operations efficiently, since these matrices are very sparse.

Following the same approach, the linear smoother can also be approximated in blockwise fashion as in~\citep{vila2017multiple}, for the predictive equations~\eqref{eq:smoother_pred1}--\eqref{eq:smoother_pred2}:
\begin{align}
    \bs_{k+1|k}^{(g)} &= \big[\bF_k\big]_{(g),:} \bs_k
    \label{eq:smoother_dist_pred1}
    \\
    \bP_{k+1|k}^{(g)} &= \big[\bF_k\big]_{(g),:} \bP_{k} \big(\big[\bF_k\big]_{(g),:}\big)^\top + \bQ_k^{(g)}
    \label{eq:smoother_dist_pred2}
\end{align}
and for the smoothing equations~\eqref{eq:smooth_up1}--\eqref{eq:smooth_up3}:
\begin{align}
    \bG_k^{(g)} & = \big[\bP_{k}\bF_k^\top\big]_{(g),(g)} \big(\bP_{k+1|k}^{(g)}\big)^{-1} 
    \nonumber \\
    & = [\bP_{k}]_{(g),(g)} \big(\big[\bF_k\big]_{(g),(g)}\big)^\top \big(\bP_{k+1|k}^{(g)}\big)^{-1} 
    \label{eq:smooth_dist_up1}
    \\
    \bs_{k|K}^{(g)} &= \bs_{k}^{(g)} + \bG_k^{(g)}\big(\bs_{k+1|K}^{(g)} - \bs_{k+1|k}^{(g)}\big)
    \label{eq:smooth_dist_up2}
    \\
    \bP_{k|K}^{(g)} &= \bP_{k}^{(g)} + \bG_k^{(g)}(\bP_{k+1|K}^{(g)} - \bP_{k+1|k}^{(g)}) (\bG_k^{(g)})^\top
    \label{eq:smooth_dist_up3}
\end{align}
where
\begin{align}
    \bG_k = \blkdiag\big\{\bG_k^{(1)},\ldots,\bG_k^{(G)}\big\} \,.
    \label{eq:smooth_dist_G}
\end{align}

One last issue is that the innovation covariance matrix $\bT_k^m$ can also be large for big images (e.g., Landsat measurements), as it has $(L_m\prod_{\ell=1}^{L_m} N_{m,\ell})^2$ elements. Fortunately, the model implicitly imposes a simple structure for this matrix. To show this, let us consider a permutation of the pixels $\bPi_m$, such that $\bPi_m \widetilde{\by}_k^m$ reorders $\widetilde{\by}_k^m$ by making different bands of each LR pixel contiguous:
\begin{align}
    \bPi_m\widetilde{\by}_k^m = 
    \begin{bmatrix}
    \begin{bmatrix}
    \widetilde{y}_{k,1,1}^{m}
    \\ \vdots \\
    \widetilde{y}_{k,L_m,1}^{m}
    \end{bmatrix}^\top 
    ,\ldots,
    \begin{bmatrix}
    \widetilde{y}_{k,1,N_m}^{m}
    \\\vdots\\
    \widetilde{y}_{k,L_m,N_m}^{m}
    \end{bmatrix}^\top 
    \end{bmatrix}^\top,
\end{align}
where $\widetilde{y}_{k,\ell,n}^{m}$ is the $n$-th pixel of the $\ell$-th band of $\widetilde{\by}_{k}$.

If we assume that $\bH_{\ell}^m$ is a local filter, i.e., each pixel in the low-resolution image is generated according to a fixed linear combination of a \textit{distinct} subset of HR pixels, this allows us to express the row-permuted version of $\widetilde{\bH}_k^m$ equivalently as:
\begin{align}
    \bPi_m\widetilde{\bH}_k^m = \blkdiag\big\{\underbrace{\bH,\bH,\ldots,\bH}_{N_m \,\rm{times}}\big\} \,,
    \label{eq:filter_conv_def}
\end{align}
where matrix $\bH\in\amsmathbb{R}^{L_m\times d^2L_H}$ is given by:
\begin{align}
    \bH = \bh^m \otimes \bC^m \,,
    \label{eq:filter_conv_def_1px}
\end{align}
where $\bC^m=\big[(\bc_1^m)^\top,\ldots,(\bc_{L_m}^m)^\top\big]^\top$ is the spectral response function for all bands, $\bh^m\in\amsmathbb{R}^{1\times d}$ is the local spatial response filter, which defined how the HI pixels inside each LR pixels are combined, and $d$ is the number of HR pixel in each LR pixel.

Using this permutation, the innovation covariance matrix can be written as:
\begin{align}
    \bPi_m\bT_k^m\bPi_m^\top 
    ={}& \bPi_m\widetilde{\bH}_k^{m}\bP_{k|k-1}\big(\widetilde{\bH}_k^{m}\big)^\top\bPi_m^\top + \bPi_m\widetilde{\bR}_k^m\bPi_m^\top
    \nonumber\\
    ={}& \blkdiag\{\bH,\ldots,\bH\}\bP_{k|k-1}\blkdiag\{\bH^\top,\ldots,\bH^\top\} 
    \nonumber\\
    & + \bPi_m\widetilde{\bR}_k^m\bPi_m^\top
    \nonumber\\
    ={}& \blkdiag\{\bH\bP_{k|k-1}^{(1)}\bH^\top, \ldots,\bH\bP_{k|k-1}^{(G)}\bH^\top\}
    \nonumber\\
    & + \bPi_m\widetilde{\bR}_k^m\bPi_m^\top \,.
\end{align}
Thus, as long as the noise is independent among different pixels (i.e., $\widetilde{\bR}_k^m$ is block diagonal), it is possible to express the innovation covariance matrix in block diagonal form by adequately permuting the LR image pixels. This shows that each pixel from the lowest resolution image modality can be processed independently when $\bQ_k$ and $\bP_{0|0}$ also have a block diagonal structure. The proposed image fusion method is summarized in Algorithm~\ref{alg:alg1}.

\begin{algorithm} [!t]
\small
\SetKwInOut{Input}{Input}
\SetKwInOut{Output}{Output}
\caption{Weakly supervised online image fusion}\label{alg:alg1}
\Input{Measured multimodal images $\by_{k}^m$, for all time instants $k=1,\ldots,K$ and modalities $m$, historical datasets of high-resolution images $\calD_k$, parameters $s_{\max}$.}

\Output{Estimated image sequence $\bs_{k|K}$}

Initialize $\bP_{0|0}$ and $\bs_{0|0}$\;

\texttt{// Filter} \;
\For{$k=1,2,\ldots,K$}{ 

Compute innovation covariance matrix $\bQ_k$ using $\bs_{k-1}$ and $\calD_k$ according to Section~\ref{sec:Qest} \;

Compute $\bs_{k|k-1}$ and $\bP_{k|k-1}$ using equations~\eqref{eq:dist_kf_groups},~\eqref{eq:dist_kf_groups_Pkold},~\eqref{eq:kalman_pred1_dist} and~\eqref{eq:kalman_pred2_dist} \tcp*{Prediction}

Compute $\bs_{k|k}$ and $\bP_{k|k}$ using equation~\eqref{eq:dist_kf_groups_Pk} and equations~\eqref{eq:kalman_dist_up1}--\eqref{eq:kalman_dist_up6} \tcp*{Update}

Constrain $\bs_{k|k}$ using~\eqref{eq:constrained_kf_simple} \;
}

\texttt{// Smoother} \;

\For{$k=K,K-1,\ldots,1$}{

Compute $\bs_{k+1|k}$ and $\bP_{k+1|k}$ using equations~\eqref{eq:dist_kf_groups},~\eqref{eq:dist_kf_groups_Pkold},~\eqref{eq:smoother_dist_pred1} and~\eqref{eq:smoother_dist_pred2} \tcp*{Prediction}

Compute $\bs_{k|K}$ and $\bP_{k|K}$ using equations~\eqref{eq:smooth_dist_up1}--\eqref{eq:smooth_dist_up3} and equation~\eqref{eq:smooth_dist_G} \tcp*{Backwards update}

}
\KwRet Estimated images $\bs_{k|K}$
\end{algorithm}

\section{Experiments}
\label{sec:experiments}

In this section, we use the proposed methodology to fuse Landsat and MODIS image over time. The Kalman filter and smoother are built under the three different assumptions for the state covariance matrices regarding the distributed implementation discussed in Section~\ref{sec:complex}:
$i)$ diagonal state covariance (denoted by KF-D and SM-D); $ii)$ block-diagonal state covariance with one block per Landsat multispectral pixel (denoted by KF-B and SM-B); and $iii)$ block-diagonal with blocks for all Landsat multispectral pixels corresponding to the same coarse pixel in a MODIS image being correlated (denoted by KF-F and SM-F). A filter in which Landsat multispectral pixels corresponding to more than one coarse pixel in a MODIS image being all correlated could not be implemented due to computational and memory limitations.

Although in our experiments we consider only two modalities the proposed methodology admits multiple different modalities provided that enough computational power is available.   
As benchmark, we compare the performance of Kalman filter and smoother under all three assumptions to that of the \emph{Enhanced Spatial and Temporal Adaptive ReFlectancefusion Model} (ESTARFM) algorithm~\citep{zhu2010fusion_ESTARFM}, and the \emph{Prediction Smooth Reflectance Fusion Model} (PSRFM)  algorithm~\citep{zhong2019improvement, zhong2018prediction}. The ESTARFM algorithm requires {two high-resolution (e.g., Landsat) images} at the beginning of the image sequence, and can generate high-resolution reconstructions at later time instants based on MODIS measurements. Thus, it is a good candidate for comparison with the Kalman filtering based strategies, which also do not require future data. The PSRFM method, on the other hand, uses two high-resolution (e.g., Landsat) images (one at the beginning and one at the end of the sequence), and provides high-resolution reconstruction for the intermediate MODIS images. Thus, it consists in an adequate comparison to the smoother algorithms, which also require future high-resolution images.
In the following, we describe the data and simulation setup, followed by the results and the discussions.


\subsection{Study region}

\begin{figure}[h!]
  \centering
\includegraphics[width=0.45\linewidth]{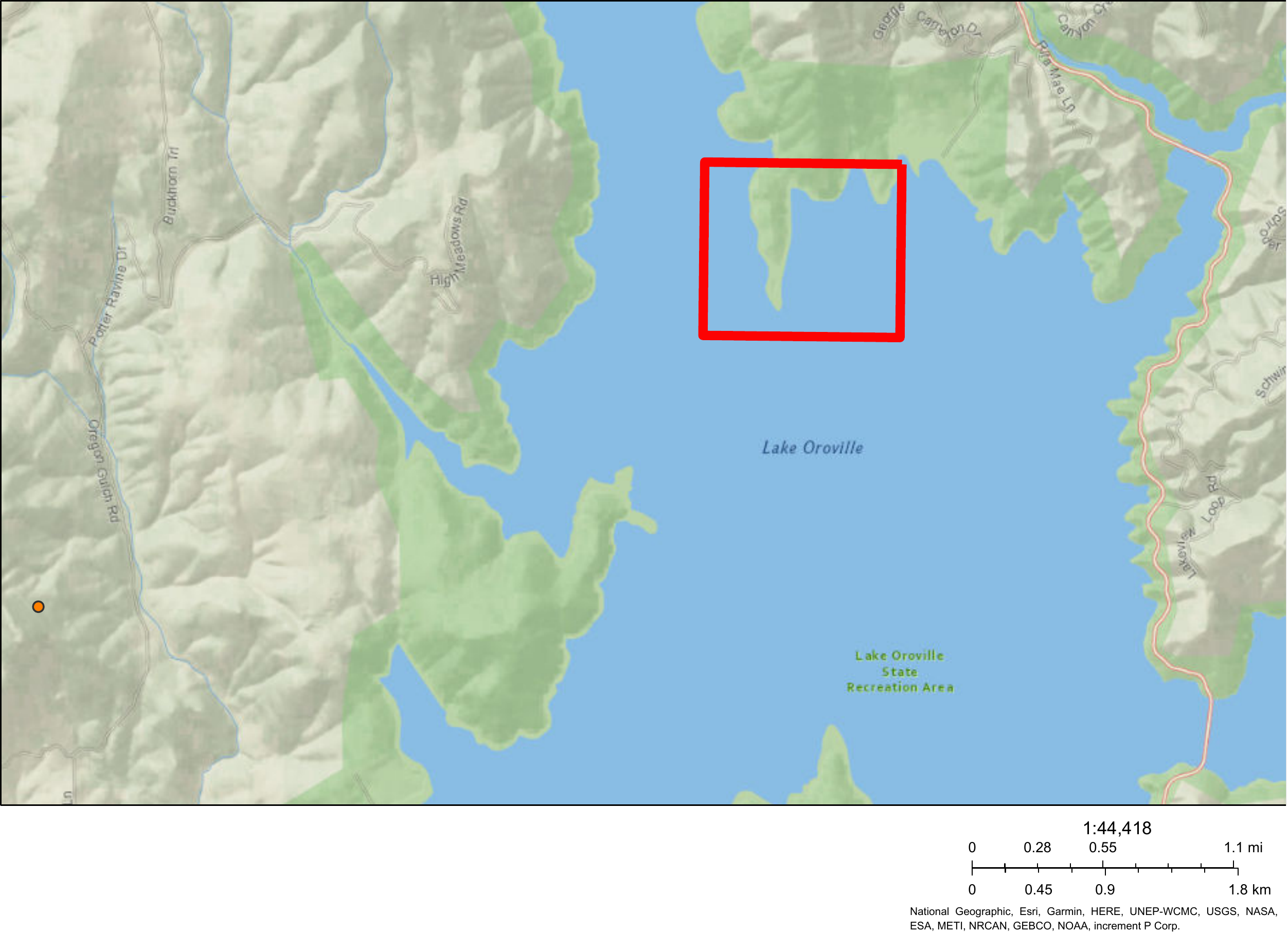} \hfil
\includegraphics[width=0.45\linewidth]{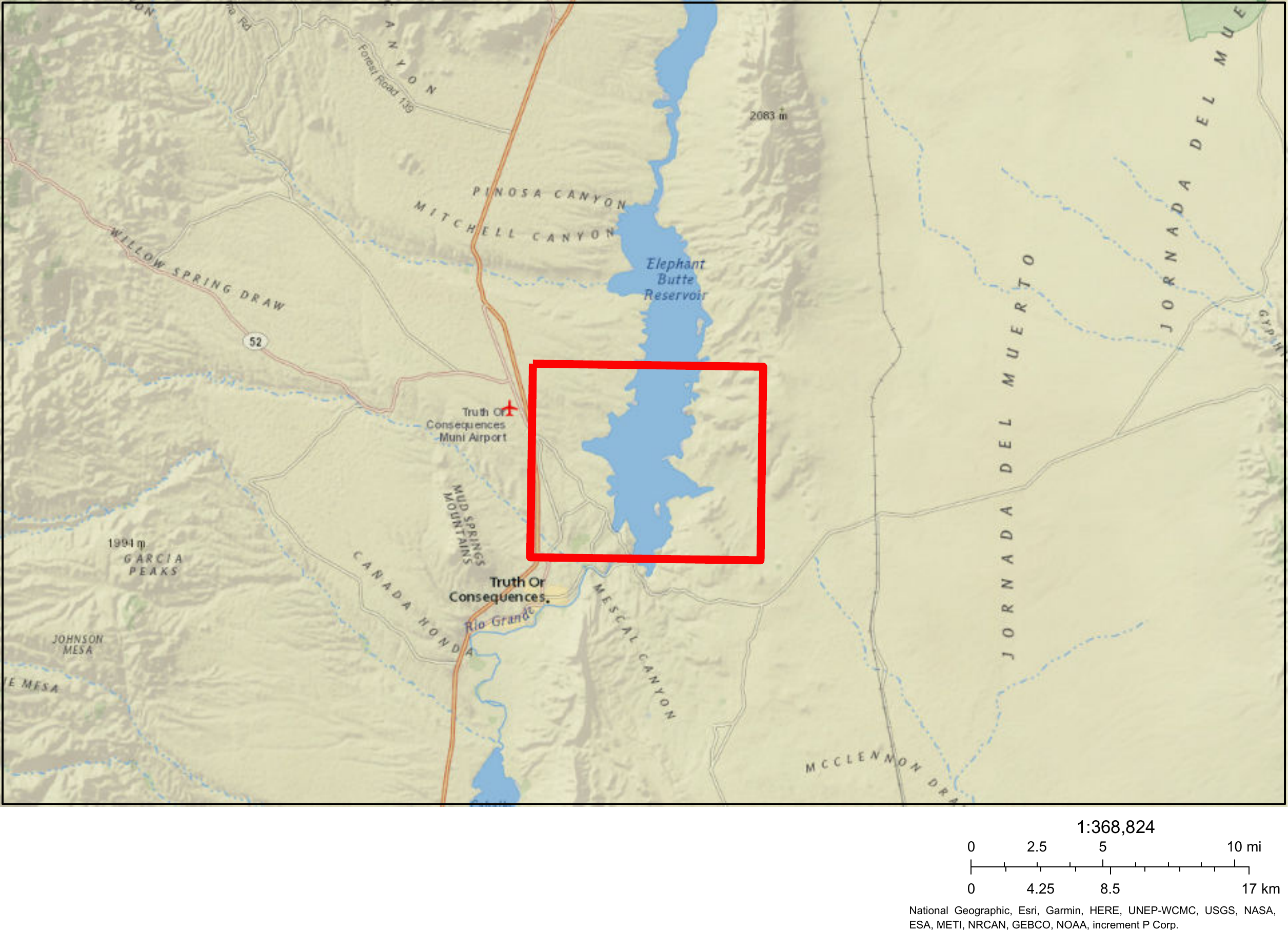}
  \caption{(Left) Oroville dam site. (Right) Elephant Butte site. The red boxes delimit the specific study areas used in our experiments.}
  \label{fig:site_location}
\end{figure}

\begin{figure}[h!]
  \centering
  \includegraphics[width=1\linewidth]{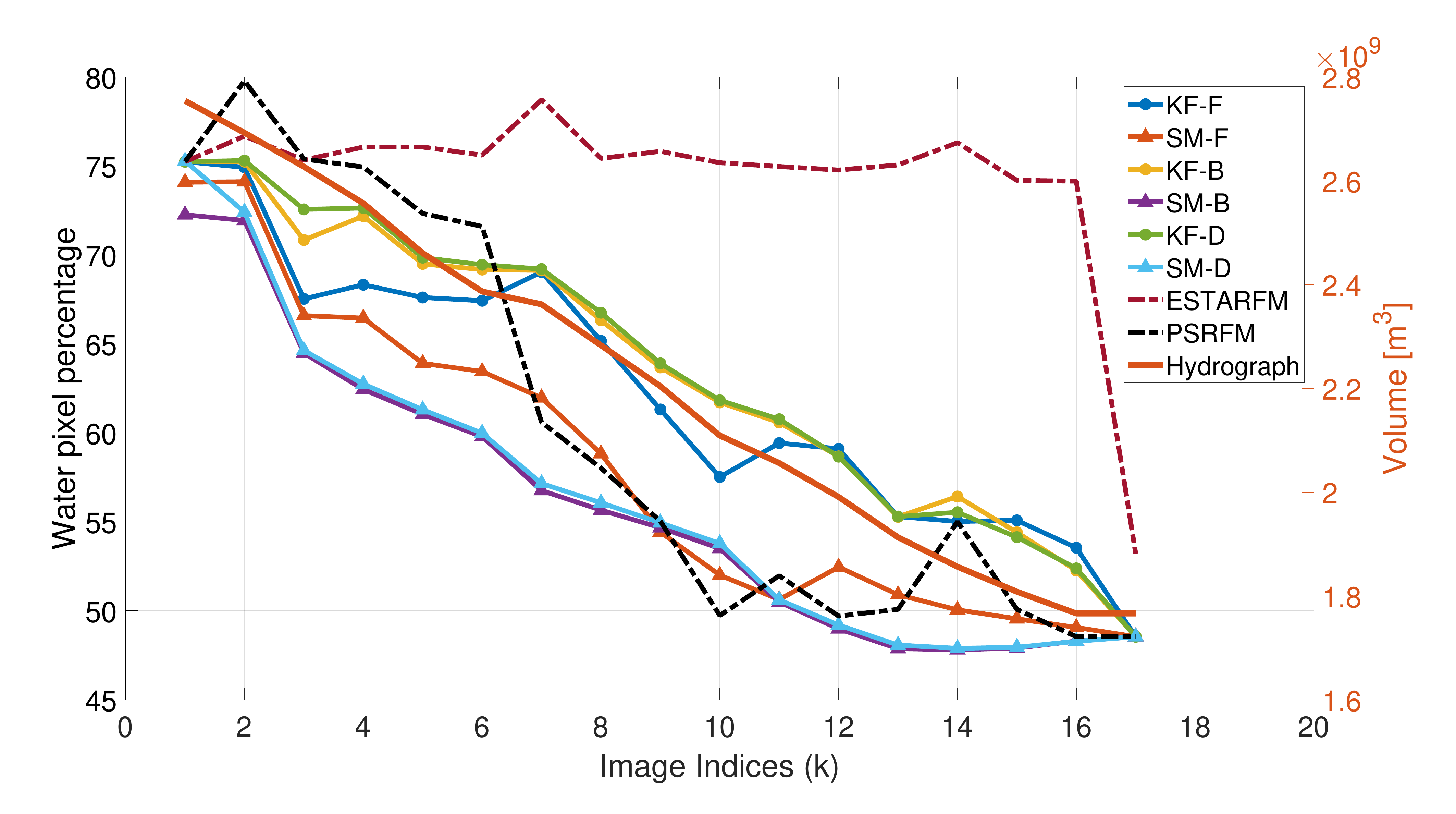}
  \vspace{-0.8cm}
  \caption{Percentage of water pixels in the estimated images over image index (time) and the reservoir volume in $m^3$ (hydrograph) for the Oroville Dam example. Classification of water was done by performing clustering on the estimated bands for each method and time index. High resolution Landsat images were observed at indices $k\in\{1, 17\}$. }
  \label{fig:hydrographest}
\end{figure}

For the experiments, we consider two sites. The first is the Oroville dam (Figure~\ref{fig:site_location}, left panel), located on the Feather River, in the Sierra Nevada Foothills (38° 35.3' North and 122° 27.8' W) is the tallest dam in USA and is major water storage facility in California State Water Project. The reservoir has a maximum storage capacity of $1.54\times10^{11}$ ft$^{3}$ or $4.36\times10^{9}$ m$^{3}$, which fills during heavy rains or large spring snow melts and water is carefully released to prevent flooding in downstream areas, mainly to prevent large flooding in Butte County and area along the Feather River. The reservoir water storage change in between 07/03 and 09/21 of 2018 is as shown as the hydrograph curve in Figure~\ref{fig:hydrographest}. Another unique characteristic is that it has three power plants at this reservoir. The water released downstream is used to maintain the Feather and Sacramento Rivers and the San Francisco-San Joaquin delta. Lake Oroville is at an elevation of 935 feet (285 meters) above sea level. 
We focus at a particular area of the Oroville dam delimited by the red box in Figure~\ref{fig:site_location}.

The second site is the Elephant Butte reservoir (Figure~\ref{fig:site_location}, right panel), located in the southern part of the Rio Grande river, in New Maxico, USA (33° 19.4' N and 107° 26.2' W). It is the largest reservoir in New Mexico, providing power and irrigation to southern New Mexico and Texas. Elephant Butte reservoir is at an elevation of 4,414 ft (1,345 meters), and has a surface area of 36,500 acres (14,800 ha).


\begin{table*} [h!]
\centering
\caption{Spectral angle mapper between the estimated high-resolution image and the Landsat measurement for the Oroville Dam example (note that the Landsat images at dates 07/19, 08/20, and 09/05 were not supplied to the algorithms and only used for evaluation purposes). However, the Landsat image at 09/21 was available to all algorithms. Note that the spectral angle is not reported for PSRFM at 09/21. This is so since PSRFM uses the last pair (MODIS-Landsat) of images and directly sets its estimations at this dates to the ground-truth.}
  \vspace{-0.2cm}
\renewcommand{\arraystretch}{1.2}
\resizebox{0.8\textwidth}{!}{%
\begin{tabular}{c||c|c|c|c|c|c|c|c}
\hline
Method & KF-F & SM-F & KF-B & SM-B & KF-D & SM-D & ESTARFM & PSRFM \\
\hline\hline
{Image (07/19)} & 7.1240  & 10.8537  & \textbf{4.2356} & 6.1515  &4.9304 & 5.9064 &6.0810 &6.8837 \\
\hline
{Image (08/20)}&  27.6343  & 26.2786  & 26.1229 &\textbf{25.1520} &27.1928 & 26.1758 &29.0892 &27.7802 \\
\hline
{Image (09/05)} & 8.5741  & 6.0366  & 6.6246 &\textbf{3.6838} &7.4482 & 4.4135  &11.4553 & 6.0354 \\
\hline
{Image (09/21)} & 8.0588  & 3.6385  & 6.4042 &\textbf{0.5471} &6.9754 & 0.6960 &11.9584 & \textbf{--} \\
\hline\hline
{Average} &  12.8478 & 11.7019  &10.8468 &\textbf{8.8836} &11.6367 & 9.2979 &14.6460 &10.1748
 \\
\hline
\end{tabular}
}
\label{tab:SAM}
\end{table*}

\begin{table*} [h!]
\centering
\caption{Percentage of misclassified pixels for the Oroville Dam example (the Landsat image at 09/21 was available to all algorithms).
Note that the misclassification percentage is not reported for PSRFM at 09/21. This is so since PSRFM uses the last pair (MODIS-Landsat) of images and directly sets its estimations at this dates to the ground-truth.} 
  \vspace{-0.2cm}
\renewcommand{\arraystretch}{1.2}
\resizebox{0.8\textwidth}{!}{%
\begin{tabular}{c||c|c|c|c|c|c|c|c}
\hline
Method & KF-F & SM-F & KF-B & SM-B & KF-D & SM-D & ESTARFM & PSRFM \\
\hline\hline
{Image (07/19)} &  9.5412   & 7.6360  &6.4472 &8.2914  & 6.1119 &8.0171 &5.4870 &\textbf{5.2431} \\
\hline
{Image (08/20)}&  14.9215 & 10.4405  & 7.8647 &4.1000 &7.2245 & \textbf{3.7799} &18.2899 &17.9851 \\
\hline
{Image (09/05)} & 13.4888  & 8.2152   & 9.6632  &4.7859 &9.4345  & \textbf{4.5877}  &22.7404 & 20.8962 \\
\hline
{Image (09/21)} &11.7360  & 3.8409  & 9.3583 &\textbf{0.2439} &9.2974 & 0.2591 &26.3374 & \textbf{--} \\
\hline\hline
{Average} &  12.4219 & 7.5332  &8.3333 &4.3553 &8.0171 & \textbf{4.1610} &18.2137 &11.0311
 \\
\hline
\end{tabular}
}
\label{tab:Misclassication}
\end{table*}

\begin{figure*}[h!]
  \centering
  \begin{subfigure}[b]{1\textwidth}
  \includegraphics[width=1\linewidth]{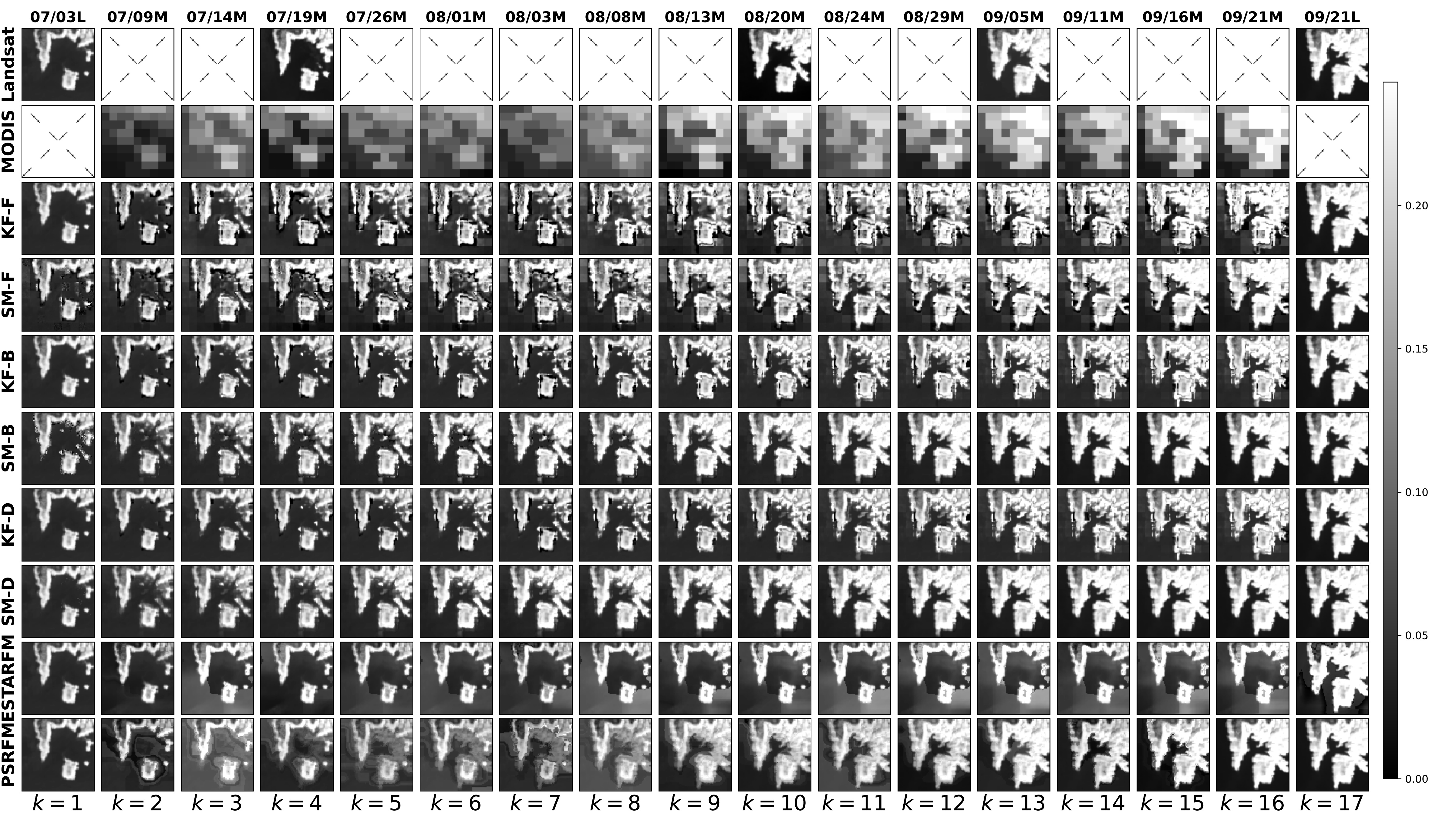}
  \vspace{-0.8cm}
  \caption{Fused images in band 1 (MODIS) and band 4 (LandSat)}
  \label{fig:Reconstruction1}
\end{subfigure}
\\\smallskip
\begin{subfigure}[b]{1\textwidth}
  \includegraphics[width=1\linewidth]{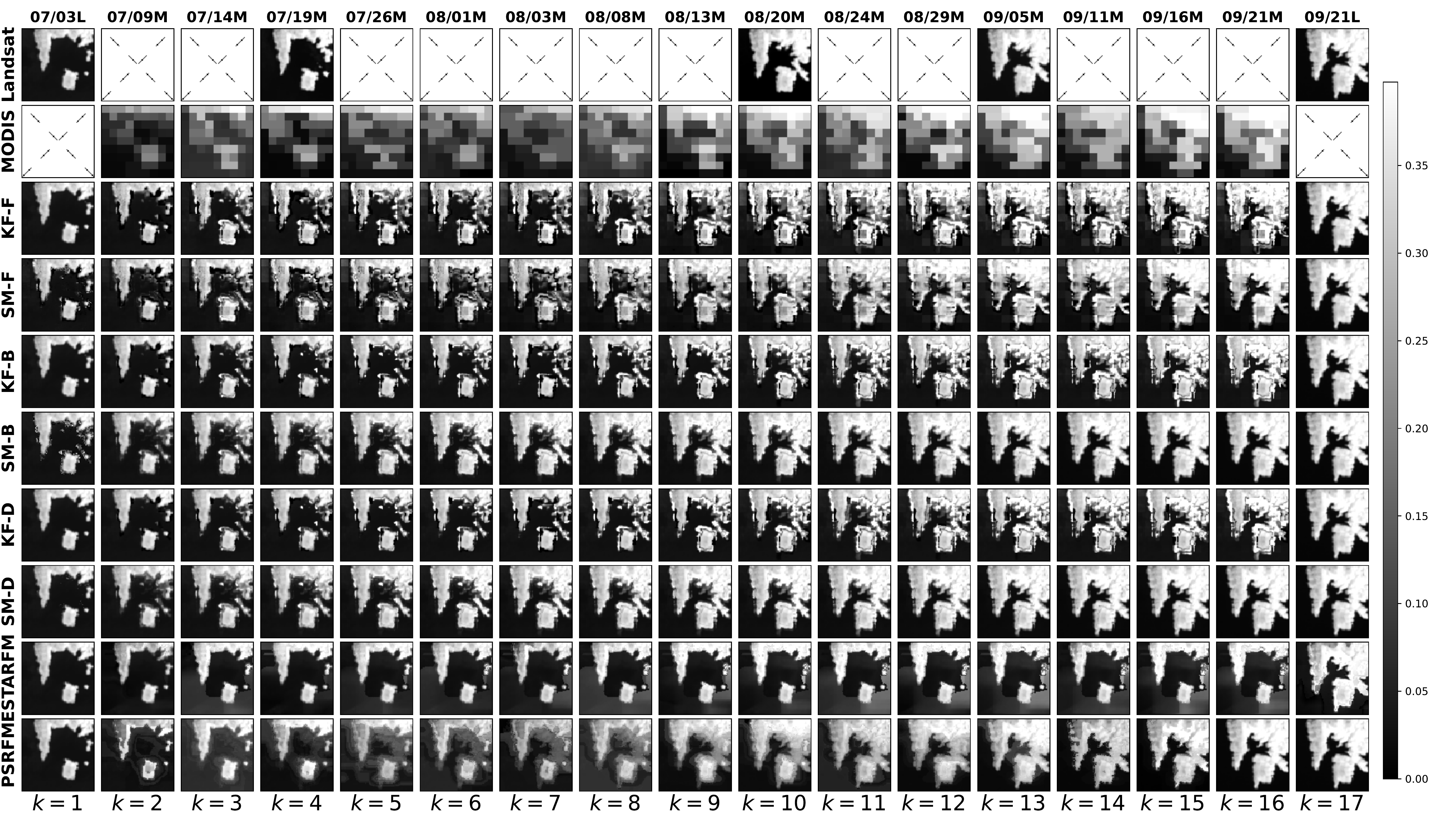}
  \vspace{-0.8cm}
  \caption{Fused images in band 2 (MODIS) and band 5 (LandSat)}
  \label{fig:Reconstruction2}
  \end{subfigure}
  \\\vspace{-0.1cm}
  \caption{Fused bands from MODIS and Landsat for the Oroville Dam example using different strategies over time. The first two rows of each subfigure depict MODIS and Landsat bands acquired at dates displayed on top labels. At each time index estimation with KF and SM under different model assumptions, ESTARFM and PSRFM are presented. Some Landsat images were omitted from the estimation process and used solely as ground-truth. Images used at each update step are indicated on top labels where ``M'' stands for MODIS and ``L'' for Landsat.}\label{fig:Reconstruction}
\end{figure*}

\begin{figure*}[h]
  \centering
\begin{subfigure}[b]{0.9\textwidth}
  \includegraphics[width=1\linewidth, height=0.8\linewidth]{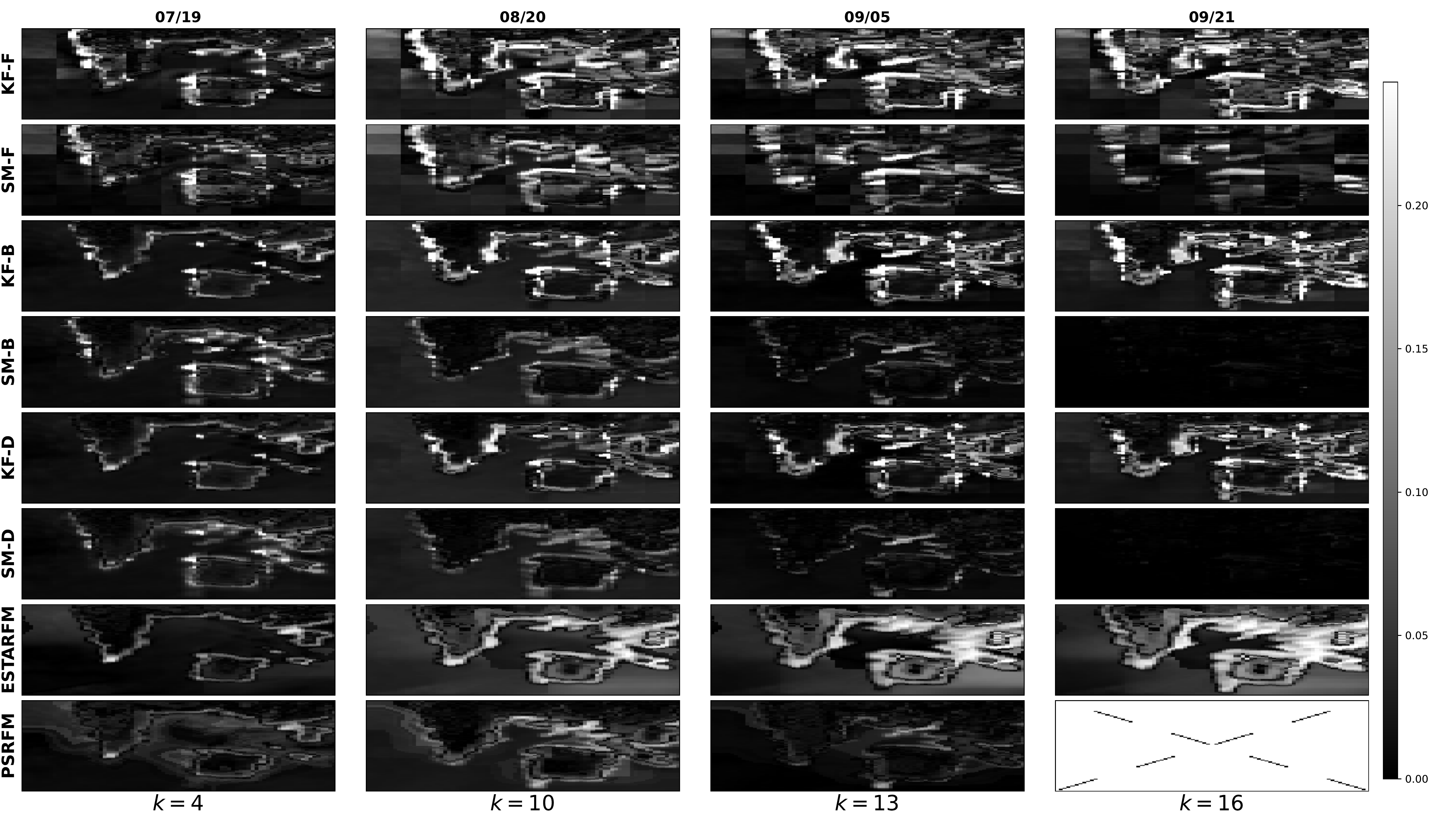}
  \end{subfigure}
%
\begin{subfigure}[b]{0.9\textwidth}
  \includegraphics[width=1\linewidth, height=0.8\linewidth]{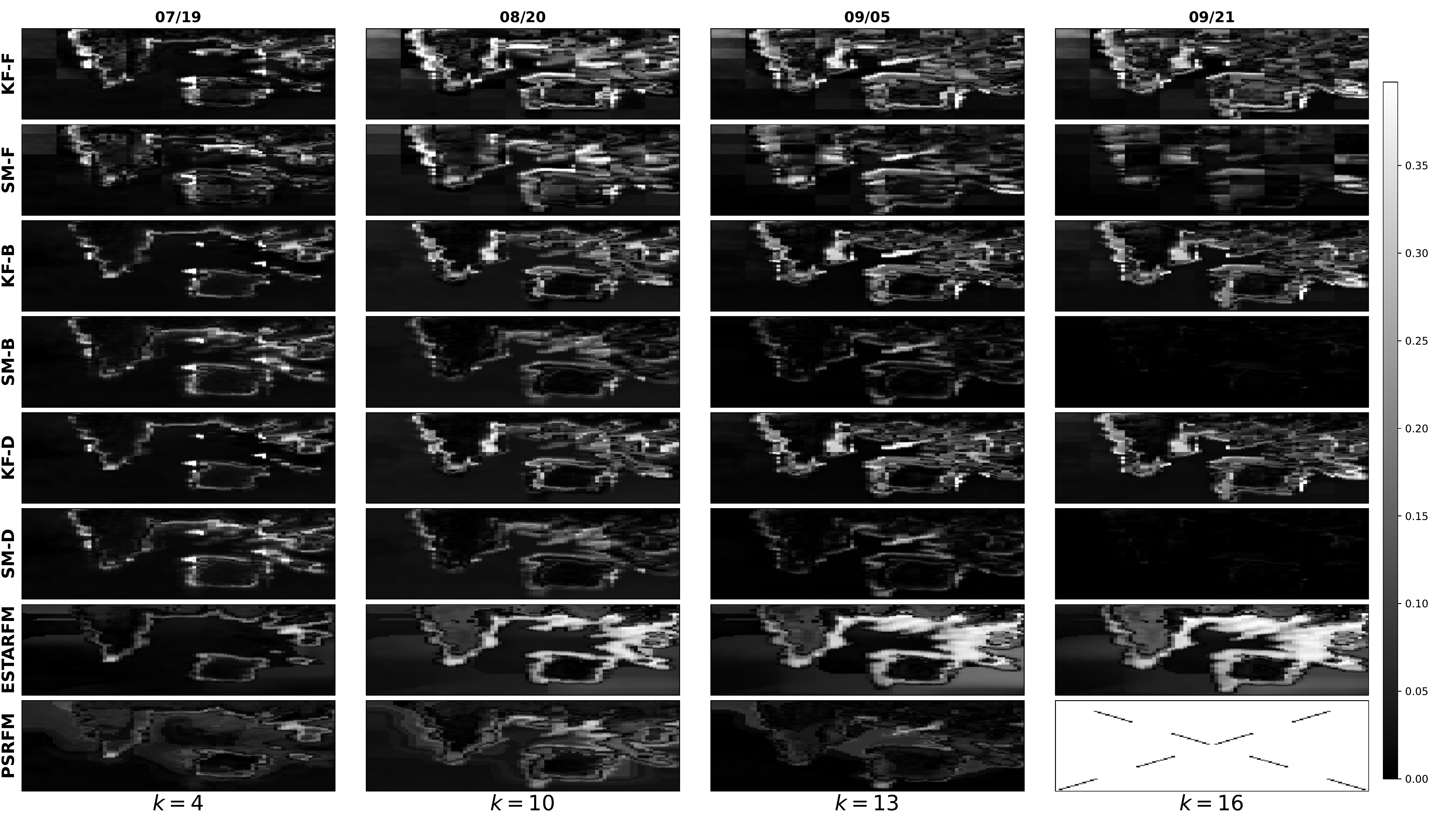}
\end{subfigure}
  \caption{Absolute difference between the estimated and ground truth (Landsat) images for the Oroville Dam example in the red (\textbf{upper panel}) and NIR (\textbf{lower panel}) bands.} \label{fig:ErrorMap}
\end{figure*}

\begin{figure*}[h]
  \centering
\begin{subfigure}[b]{1\linewidth}
  \centering
  \includegraphics[width=1\linewidth]{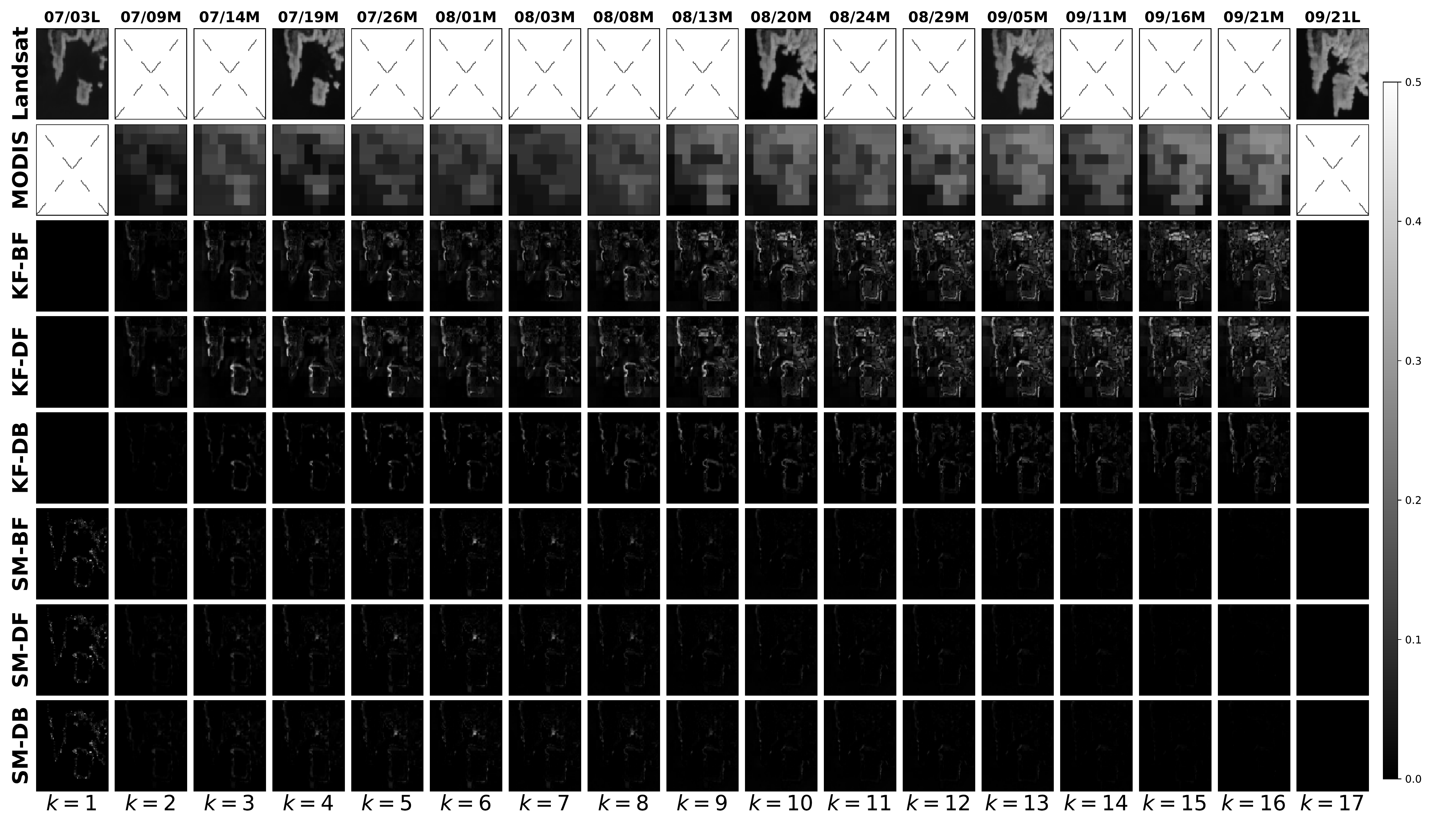}
  \label{fig:errormap1}
  \end{subfigure} 
%
\vspace{-0.6cm}
\begin{subfigure}[b]{1\linewidth}
  \centering
  \includegraphics[width=1\linewidth]{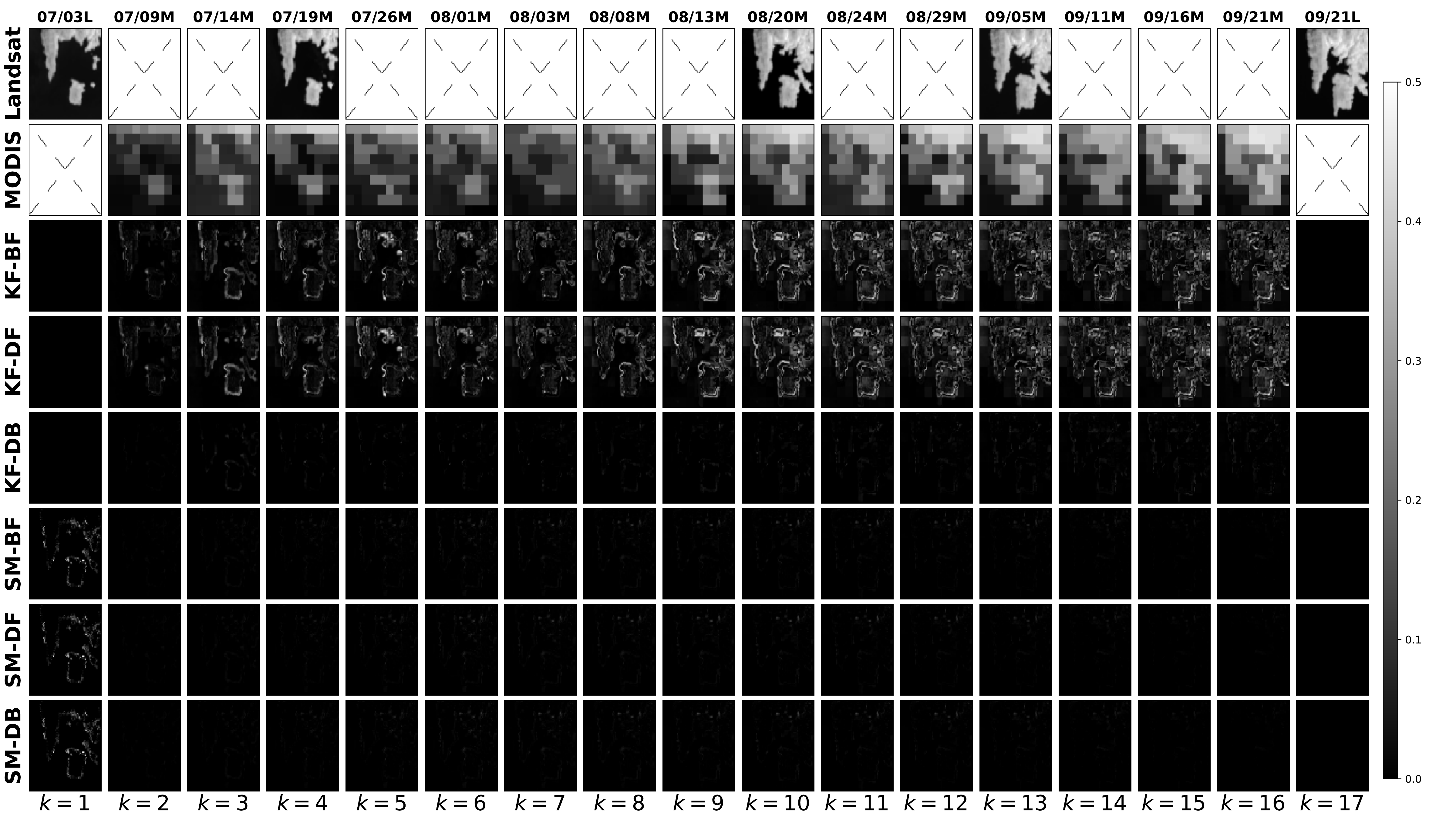}
  \label{fig:errormap2}
  \end{subfigure}
  \caption{Absolute differences between the images estimated by the KF and Smoother under different model assumptions for red (\textbf{upper panel}) and NIR (\textbf{lower panel}) bands, for the Oroville Dam example. KF-BF: difference between the estimates of KF-B and KF-F. KF-DF: difference between the estimates of KF-D and KF-F. KF-DB: difference between the estimates of KF-D and KF-B. An analogous notation holds for the smoother (SM) estimates.}
  \label{fig:DifferenceMap}
\end{figure*}

\begin{figure*}[h]
  \centering
  \begin{subfigure}[b]{0.88\textwidth}
  \includegraphics[width=1\linewidth, height=0.8\linewidth]{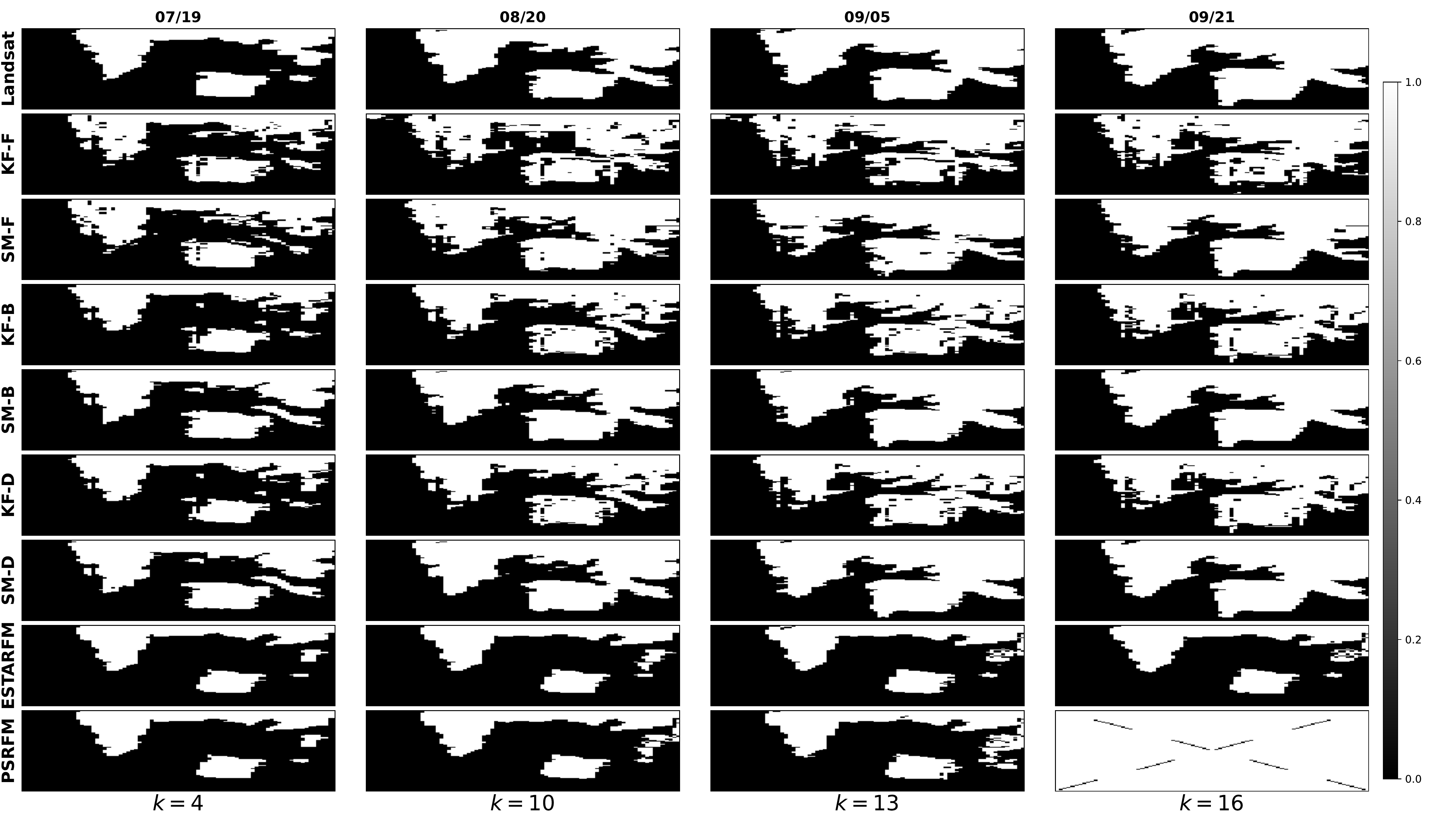}
  \end{subfigure}
\begin{subfigure}[b]{0.88\textwidth}
  \centering
  \includegraphics[width=1\linewidth, height=0.8\linewidth]{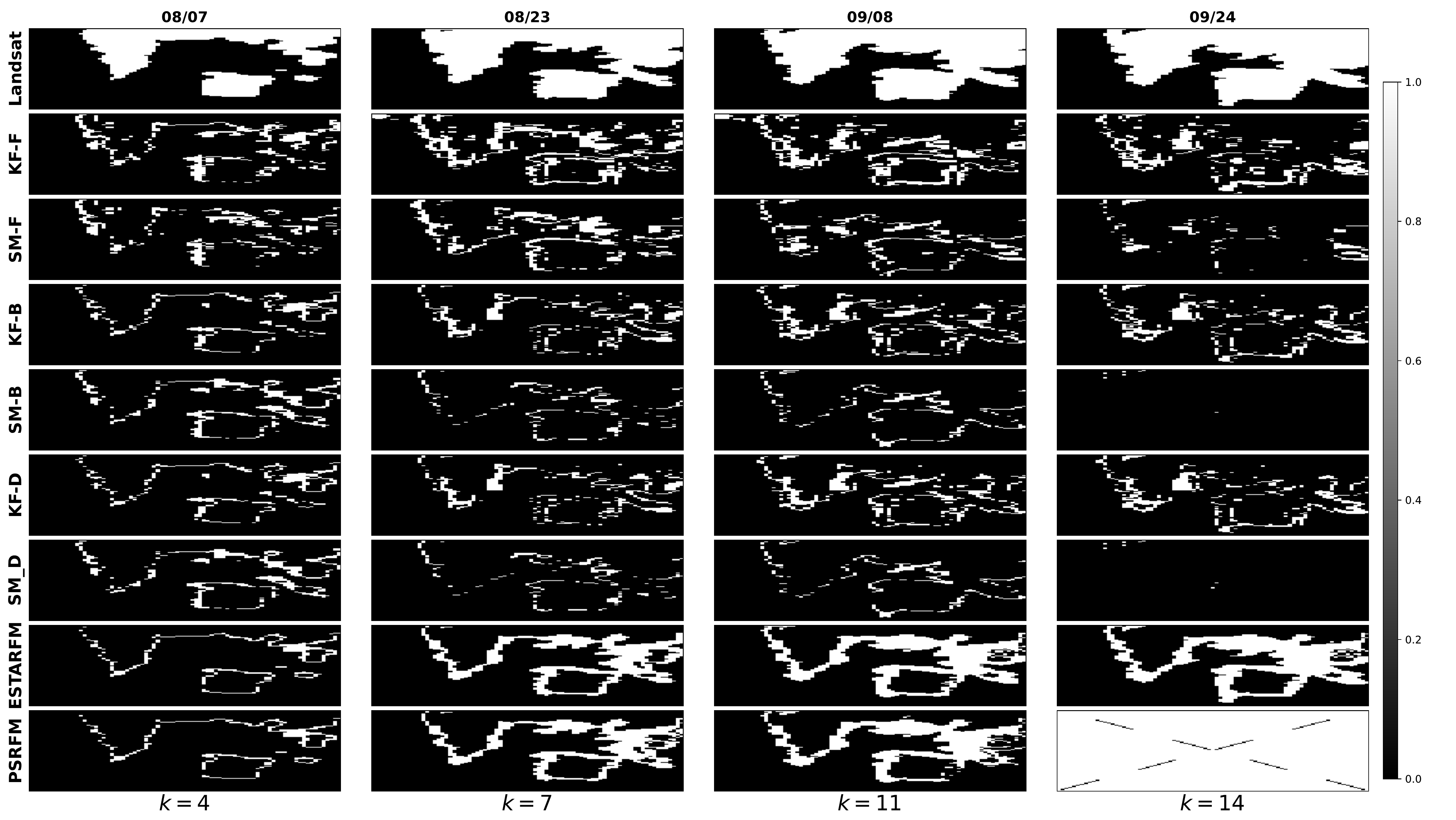}
\end{subfigure}
\vspace{-0.4cm}
\caption{(\textbf{Upper Panel}) Water map of the reconstructed images of the Oroville Dam example based on K-means clustering strategy, where 1 indicates land and 0 indicates water pixels. Classification maps obtained from Landsat images not observed by the image fusion algorithms establish the ground-truth (first row). (\textbf{Lower Panel}) Absolute error of Water map of images based on K-means clustering strategy, where 0 indicates correctly classified pixels and 1 indicates misclassifications. The ground-truth is shown in the first row.} 
\label{fig:watermap}
\end{figure*}

\begin{figure*}[h]
  \centering
\begin{subfigure}[b]{0.8\textwidth}
  \centering
  \includegraphics[width=1\linewidth]{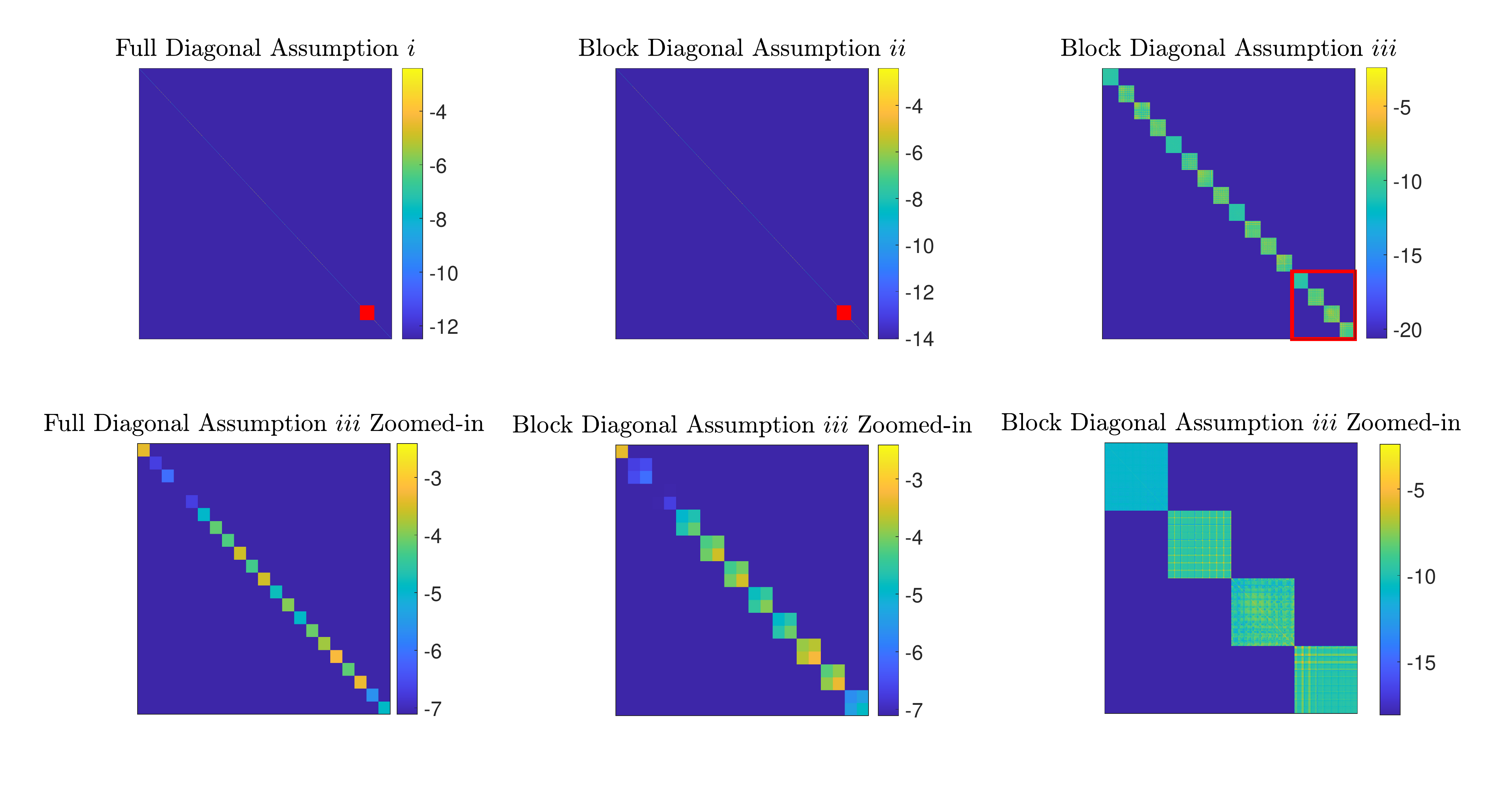}
\end{subfigure}
%
\\
\begin{subfigure}[b]{0.6\textwidth}
  \centering
  \includegraphics[width=1\linewidth]{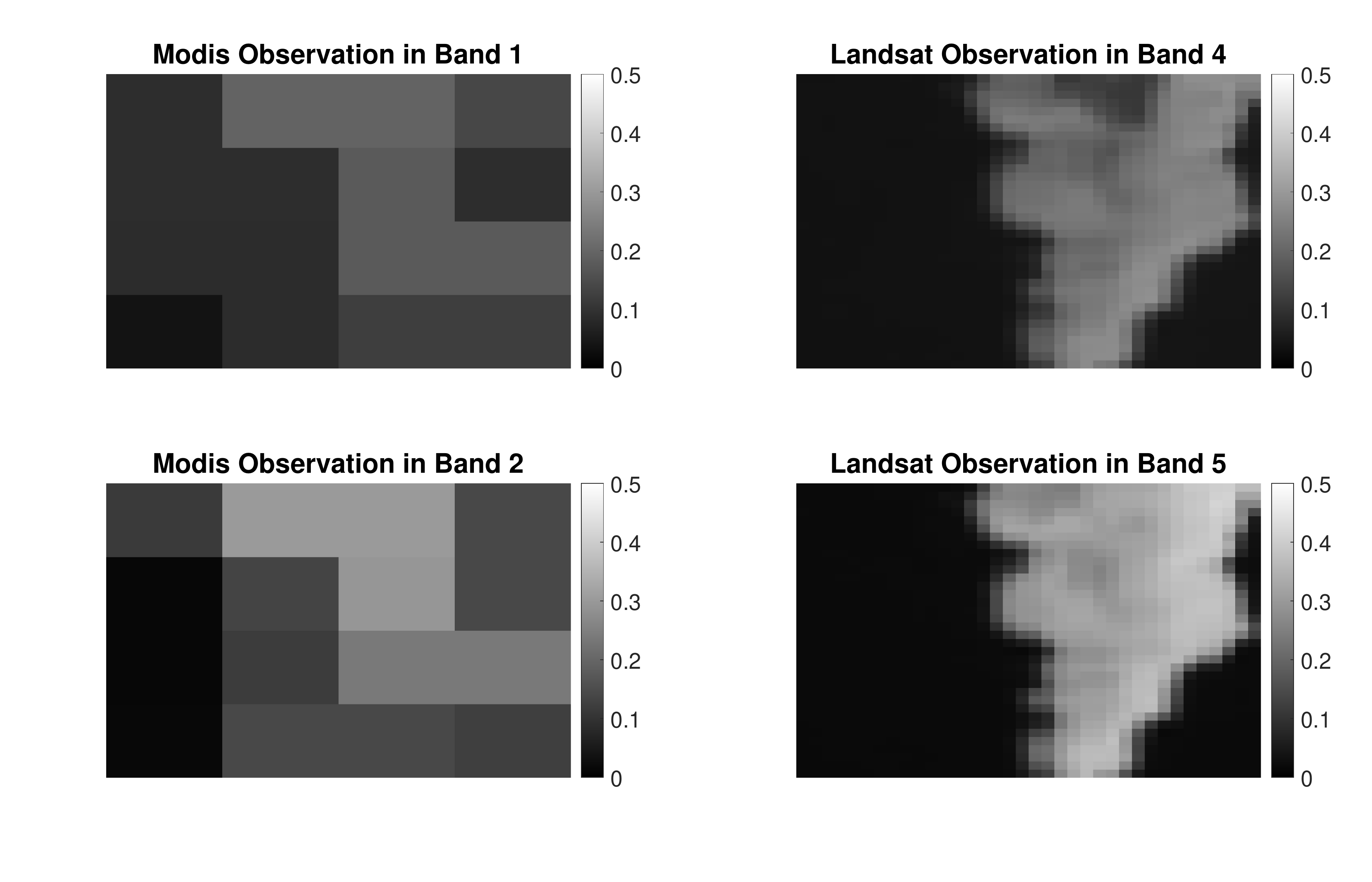}
\end{subfigure}
\vspace{-0.5cm}
  \caption{(\textbf{Top Colored Panel}) Estimated state covariance structure of the Kalman filter under model assumptions $i$, $ii$ and $iii$ for a small image area in the Oroville Dam example and $k=13$. Top row depicts the whole covariance matrix with a red square indicating the zoomed part displayed on the bottom row. The plots indicate that correlations are present when assuming block diagonal covariance matrices. (\textbf{Bottom Panel}) Zoom of the MODIS image for bands~1 and~2 (left), and the corresponding Landsat observations for bands~4 and~5 (right) corresponding to the covariance matrices plotted in the right panels.} \label{fig:covstructobserv}
\end{figure*}


\subsection{Remote Sensed data}


For our simulations with the Oroville Dam site, we collected MODIS and Landsat data acquired from the region marked with a red square on Figure~\ref{fig:site_location}, and on a interval ranging from $2018/07/03$ to $2018/09/21$. This interval was selected since the hydrograph analysis indicates high variation in the water level of the reservoir, see, the hydrograph curve in Figure~\ref{fig:hydrographest}. Such variation in the water levels result in large changes in the acquired images, exposing flooded areas. In this experiment we will focus on the red and near-infrared (NIR) bands since they are often used to distinguish water from other landcover elements in the image~\citep{gao1996ndwi}. We also collected $5$ Landsat data from $2017/08/01$ to $2017/12/07$ to serve as a past historical dataset~$\cp{D}_k$.

\interfootnotelinepenalty=10000

The study region marked in the left panel of Figure~\ref{fig:site_location} corresponds to Landsat and MODIS images with $81\times 81$ and $9\times 9$ pixels, respectively\footnote{The Landsat images were also upsampled to a spatial resolution of 27.77 meters to make its resolution exactly 9 times that of MODIS.}. After filtering for heavy cloud cover during the designated time periods, a set of 6 Landsat and 16 MODIS images were obtained. We used the first MODIS and Landsat images for initialization of all methods leading to 5 and 15 images used in the remaining fusion process. 


From the set of 5 Landsat images of the Oroville Dam site that were available for testing, three of them were set aside and not processed by any of the the algorithms. These images were acquired at dates 07/19, 08/20 and 09/05, when MODIS observations were also available, and will be used in the form of a reference for the evaluation of the algorithms' capability of estimating the high resolution images at these dates solely from the low resolution MODIS measurements.




For the simulations with the Elephant Butte site, shown in the right panel of Figure~\ref{fig:site_location}, we aim to evaluate the performance of the algorithms when processing a larger geographical area, with an area of approximately $9km \times 9km$. The setup is similar to the Oroville Dam example. We focus on the red and near-infrared bands of the Landsat and MODIS instruments, and collect 47 Landsat images from 2014/01/16 to 2017/11/24 to serve as the past historical dataset~$\cp{D}_k$.

The study region corresponds to Landsat and MODIS images with $324\times 324$ and $36 \times 36$ pixels, respectively. After removing images with significant cloud cover, we obtained a set of 5 Landsat and 7 MODIS images to process. We used the first MODIS and Landsat image pair to initialize the algorithms, leading to 4 Landsat and 6 MODIS images to be used in the remaining fusion process. From the set of 4 Landsat images that were available for testing, 2 of them were set aside as ground truth to evaluate the algorithms. Theses images are acquired at dates 06/07 and 06/23. However, the MODIS measurements at those dates contained significant cloud cover, and had to be discarded. Therefore, we evaluate the performance of the algorithms through the estimation results obtained dates 06/14 and 06/27 (in which the MODIS observations were available).


\begin{figure*}[h!]
  \centering
  \begin{subfigure}[b]{1\textwidth}
  \centering
  \includegraphics[width=0.95\linewidth]{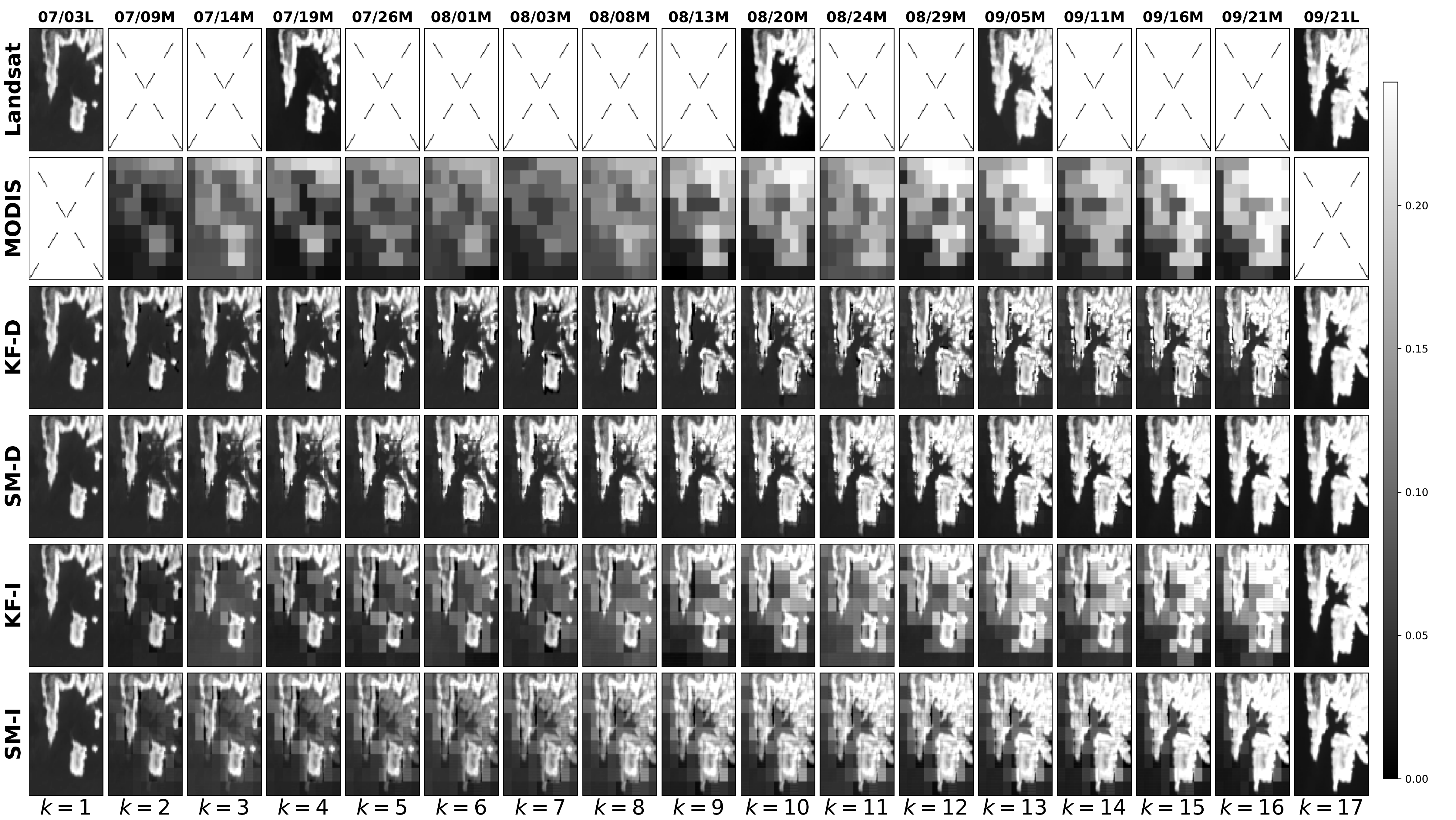}
  \vspace{-0.15cm}
  \caption{Fused images in band 1 (MODIS) and band 4 (LandSat)}
  \label{fig:KF_KFQ1}
\end{subfigure}
\\\smallskip
\begin{subfigure}[b]{1\textwidth}
  \centering
  \includegraphics[width=0.95\linewidth]{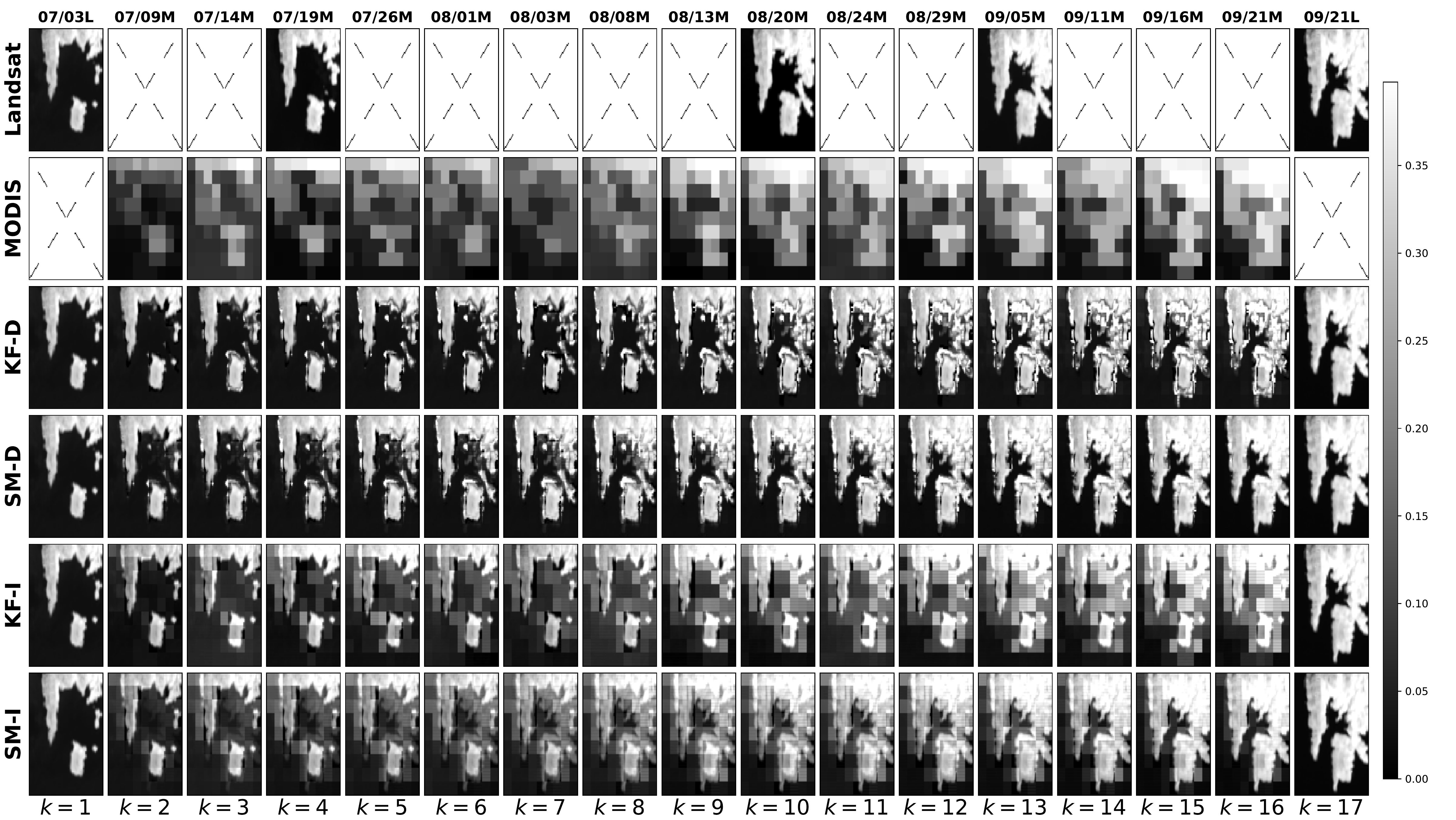}
  \vspace{-0.15cm}
  \caption{Fused images in band 2 (MODIS) and band 5 (LandSat)}
  \label{fig:KF_KFQ2}
  \end{subfigure}
  \\\vspace{-0.2cm}
  \caption{Fused bands from MODIS and Landsat for the Oroville Dam example using different strategies over time. The first two rows of each subfigure depict MODIS and Landsat bands acquired at dates displayed on top labels. At each time index estimation results of the diagonal Kalman filter and smoother with the proposed weakly supervised calibration strategy (KF-D and SM-D) are compared to the result of a Kalman filter and smoother with $\bQ_k$ being proportional to the identity (denoted by KF-I and SM-I). Landsat images at dates 07/19, 08/20 and 08/29 were omitted from the estimation process and used solely as ground-truth. Images used at each update step are indicated on top labels where ``M'' stands for MODIS and ``L'' for Landsat.}\label{fig:KF_KFQ}
\end{figure*}

\subsection{Algorithm setup}

We initialized the proposed Kalman filter and smoother using a high resolution Landsat observation as the state, i.e., $\bs_{0|0}=\widetilde{\by}_0^{\mathsf{L}}$, and set $\bP_{0|0}=10^{-10}\bP_0$. The structure of $\bP_0$ varies with different assumptions: $i)$ $\bP_0 = \bI$ if the state covariance is diagonal; $ii)$ $\bP_0 = \text{blkdiag}\{\bP_{0,1}, \bP_{0,2}, \cdots, \bP_{0, N_H}\}$, where
%
$\bP_{0,i} = \frac{1}{2}\mathbb{1}+\frac{1}{2}\bI$, with $\mathbb{1}$ being an all ones matrix, if the state covariance matrix has a block-diagonal structure with one block per Landsat multispectral pixel; $iii)$ $\bP_0 = \text{blkdiag}\{\bP_{0,1}, \bP_{0,2}, \cdots, \bP_{0, \tilde{N}_m\times L_m}\}$, where
%
$\bP_{0,i} = \frac{1}{2}\mathbb{1}+\frac{1}{2}\bI$
if the state covariance matrix has a block-diagonal structure with each block containing all Landsat multispectral pixels corresponding to the same coarse pixel in a MODIS image. Figure~\ref{fig:covstructobserv} shows an example of the final $\bP_{k|k}$, {$k=13$}, obtained with the KF under all the assumptions discussed in Section~\ref{sec:complex}. The noise covariance matrices were set as $\bR_{\ell}^{\mathsf{L}}=10^{-10}\bI$ and $\bR_{\ell}^{\mathsf{M}}=10^{-4}\bI$, for all $\ell$. The blurring and downsampling matrices were set as $\bH_{\ell}^{\mathsf{L}}=\bI$ for Landsat, while for MODIS $\bH_{\ell}^{\mathsf{M}}$ consisted of a convolution by an uniform $9\times9$ filter, defined by $\bh=\frac{1}{81}{\mathbb{1}_{9\times9}}$ (where ${\mathbb{1}_{9\times9}}$ is a $9\times9$ matrix of ones), followed by decimation by a factor of~$9$, which represents the degradation occurring at the sensor (see, e.g.,~\citep{huang2013spatiotemporalFusionBayes}). We also set $\bF_k=\bI$ for all $k$. The vectors $\bc_{\ell}^m$ contained a positive gain in the $\ell$-th position which compensated for scaling differences between Landsat and MODIS sensors, and zeros elsewhere.

The matrices $\bD_k^{\mathsf{M}}$ were constructed based on the quality codes (i.e., the QA bits) released by MODIS for each image pixel~\citep{Vermote_modis_2015}. 
QA bits provides information regarding pixel quality and cloud cover for all pixels and all bands. In our experiments we dropped any pixel not classified as \emph{corrected product produced at ideal quality} in the QA bits~\citep{Vermote_modis_2015} by adding zeros at corresponding positions in $\bD_k^{\mathsf{M}}$.
Matrices $\bQ_k$ were computed following our data-driven strategy described in Section~\ref{sec:Qest} where $\varepsilon^2=10^{-5}$ and $n=1$.

The ESTARFM algorithm was parametrized as follows \citep{zhu2010fusion_ESTARFM}, $w = 14$ as half of the window size, the number of classes was set to $4$, and the pixels range was set to $[0, 0.5]$.
The PSRFM algorithm was parametrized as follows, $\texttt{CLUSTER}\_\texttt{METHOD} = \texttt{KMEAN}$, and $\texttt{CLUSTER}\_\texttt{DATA} = \texttt{fine}+\texttt{coarse}$. 
We highlight that all methods have access only to the first (07/03) and last (09/21) Landsat images, which allows the algorithms to produce estimates for the MODIS images observed from the second (07/09, $k=2$) up to the last date (09/21, $k=16$). However, PSRFM uses the the last pair (MODIS-Landsat) during its inference process. For this reason, error metrics computed for PSRFM on (09/21) should be disregarded as the estimate is directly the ground-truth (i.e., the Landsat image) and, thus, are not reported in the experimental results.

All algorithms are evaluated using three metrics, which are computed taking as reference the Landsat images, three of which are not observed by the algorithms. The first metric is the Spectral Angle Mapper (SAM), which attempts to measure the estimation accuracy directly:
\begin{align}
    \operatorname{SAM}(\bS,\widehat{\bS}) = \frac{1}{N_H}\sum^{N_H}_{r=1}\arccos\Big(\frac{\bs_r^\top\widehat{\bs}_r}{\|\bs_r\| \|\widehat{\bs}_r\|}\Big) \,,
\end{align}
where $\bS$ and $\widehat{\bS}$ denote the true and the estimated images, respectively. $\bs_r$ and $\widehat{\bs}_r$ denote the $r$-th pixels of different bands in $\bS$ and $\widehat{\bS}$, respectively. The two remaining metrics are related to downstream tasks of water classification and water level monitoring, which are performed on the reconstructed image sequence. 

We evaluate the direct benefit of the different fusion strategies in classifying water pixels from the estimated images. To classify water pixels we resorted to a KNN classifier whose centroids of water and non-water 2-band pixels were computed using K-Means algorithm. 
Finally, we evaluate the performance of the algorithms for hydrograph estimation by plotting the proportion of pixels in the image classified as water over time against the true hydrograph for the period, for all algorithms.





\subsection{Results for the Oroville Dam site}

As discussed, we fused the red and NIR reflectance bands of MODIS and Landsat for the selected study region. In Figure~\ref{fig:Reconstruction}, we show the fused red (Figure~\ref{fig:Reconstruction1}) and NIR (Figure~\ref{fig:Reconstruction2}) reflectances as well as the acquired red and NIR reflectance values from MODIS and Landsat. Acquisition dates are displayed in the top labels at each column with a character, $M$ for MODIS and $L$ for Landsat, indicating the image used in the fusion algorithms. We recall that only the first and last Landsat images were used in the fusion process, keeping the remaining three images as ground-truth for evaluation purposes. 
Analyzing the results we can see that the images estimated by the proposed Kalman filter and smoother methods, under different assumptions, produce better visual similarity with the Landsat (ground-truth) images for both bands. For instance, the increase in the island and the expansion of other land parts are clearly visible for the proposed methods. In contrast, analyzing ESTARFM results we note that land parts remain mainly constant through time until a new Landsat image is observed. Although lighter areas on the water portions can be noticed, specially for $k>8$, its distribution does not resemble the ground-truth. This is expected since ESTARFM is not designed to acknowledge prior information or historical data. PSRFM results show an improvement compared with ESTARFM results, since it uses both the first and the last Landsat images. However, the PSRFM results does not resemble the ground-truth very closely, and significant blurring occurs around the edge of the island when~$k<8$.
The blurring results in PSRFM are caused by the fact that the reconstructions provided by this algorithm are based on a form of interpolation which does not consider any information about the transition of the pixel reflectance values, whereas in our proposed methods we use the historical data to calibrate the time-varying dynamical model by means of matrix $\bQ_k$, which can increase the accuracy of the estimations.

Note that the images estimated by KF-F and SM-F (which used a full state covariance matrix) contained more artifacts when compared to the ones obtained by KF-B, SM-B, KF-D and SM-D (which constrained the state covariance matrix to be diagonal or block diagonal). This occurs due to the high-dimensionality of the state vector (i.e., equivalent to a vectorized Landsat image) when compared to the MODIS measurements, as this leads to the amount of measurements not being sufficient to provide an accurate estimate of the full state vector and its covariance matrix, as shown in~\cite{furrer2007estimation}. Thus, the extra degrees of freedom of KF-F and SM-F end up impacting their performance negatively. By setting the covariance matrix of the Kalman filter and smoother to be block diagonal or fully diagonal, the amount of parameters to be estimated is greatly reduced in KF-B, SM-B, KF-D and SM-D, leading to better results.

The results discussed above are corroborated by the absolute error maps displayed in Figure~\ref{fig:ErrorMap}, and SAM results shown in Table~\ref{tab:SAM} for dates in which ground-truth is available.
Analyzing Figure~\ref{fig:ErrorMap} we highlight that SM-B and SM-D clearly present the smallest errors (i.e., overall darker pixels) for both bands and all dates. KF-B also presents low absolute error except for contour regions. PSRFM is the third overall darker image, followed by KF-B, KF-D, SM-F, KF-F and ESTARFM with exception of the results on 07/19 (first column), where ESTARFM is close to the ground-truth. Similar conclusions can be achieved by analyzing Table~\ref{tab:SAM}.
The difference between the images estimated by the Kalman filter and smoother under the different approximations for the state covariance matrices (which are discussed in Section~\ref{sec:complex} and illustrated for this example in Figure~\ref{fig:covstructobserv}) is shown in Figure~\ref{fig:DifferenceMap}. It can be seen that the approximations had a more pronounced effect on the Kalman filter compared to the smoother. Moreover, the differences between the filter with a diagonal (assumption $i$) and block diagonal state covariance with one block per Landsat pixel (assumption $ii$) was relatively small. Taking in to consideration the quantitative metrics in Table~\ref{tab:SAM}, this indicates that using a diagonal or block diagonal assumption on the state covariance matrix with small blocks has a positive effect on the estimation performance, which likely occurs since it drastically reduces the amount of unknowns in the model that have to be estimated by the methods.

The left panel in Figure~\ref{fig:watermap} presents the water maps for the ground-truth (first row) and all studied algorithms obtained using K-means clustering, while the right panel in  Figure~\ref{fig:watermap} shows the misclassification maps (i.e., the absolute error between the water maps obtained by each algorithm and the ground-truth). When comparing the resulting classification maps and the misclassification error with the ground-truth, the proposed methods present classification maps that are semantically better than the competing methods.
This conclusion is also reached by considering the quantitative misclassification results presented in Table~\ref{tab:Misclassication}, in which the Kalman filter- and smoother-based methods led to smaller misclassification rates for all images except the ones on 07/19 and 09/21. 
A closer analysis reveals that the SM-D and SM-B methods hold the first and second best performance on average, followed by SM-F, KF-D, KF-B, PSRFM, KF-F and ESTARFM. Note that the PSRFM method requires access to the ground-truth (Landsat image) on 09/21 in order to produce an estimation for the MODIS image observed in this same date (i.e., measurement $k=16$), which is why the corresponding misclassification percentage is not reported. We also remark that KF-D and KF-B also obtained competitive misclassification performance (i.e., better than PSRFM), despite using no knowledge of the Landsat image at 09/21.
Moreover, comparing the results in Table~\ref{tab:SAM} and~\ref{tab:Misclassication}, it can be seen that the higher SAM results observed for all methods at date 08/20 does not translates into a worse classification performance. This indicates that the SAM results at this date were influenced by the acquisition conditions of the Landsat image which was used for ground truth, making the classification performance more straightforward to interpret.


Finally, we plotted the percentage of pixels classified as water over the time index $k$ in Figure~\ref{fig:hydrographest}, as well as a hydrograph which serves as an indicative of the dynamical evolution of the true level of the reservoir over time. It can be seen that ESTARFM was not able to properly identify the dynamical evolution of the reservoir level, leading to an estimation that was almost constant for all~$k<17$ and very different from the hydrograph curve. PSRFM led to results that, although showing relatively high day-to-day variations, were closer to the hydrograph curve. The Kalman filter and smoother-based algorithms, particularly those with the diagonal and block diagonal state covariance assumption (KF-D, KF-B, SM-B and SM-D) led to curves that were very close to the hydrograph. Thus, the Kalman filter methods captured the general trends of the hydrograph curves, even without having access to information from the Landsat image at the end of the sequence (like the smoothers and PSRFM). We note, however, that the connection between the hydrograph and the water surface area is indirect; thus, small differences between the algorithms have to be interpreted with proper care.


\subsection{Contribution of the temporal dynamics calibration strategy}

This subsection aims to show the impact of the proposed calibration strategy, which learns the temporal dynamical model parameters $\bQ_k$ using historical data, on the performance of the proposed KF and SM algorithms. To this end, we compared the proposed KF-D and SM-D (which estimate $\bQ_k$ and use a diagonal assumption on the state covariance matrix), to
a Kalman filter and smoother with a fixed $\bQ_k=10^{-2}\bI$, which we denote by KF-I and SM-I, respectively.
In Figure~\ref{fig:KF_KFQ}, we show the fused red (Figure~\ref{fig:KF_KFQ1}) and NIR (Figure~\ref{fig:KF_KFQ2}) reflectance images, as well as the acquired red and NIR reflectance values from MODIS and Landsat. Acquisition dates are displayed in the top labels at each column with a character, $M$ for MODIS and $L$ for Landsat indicating the image used in the fusion algorithms. We recall that only the first and last Landsat images were used in the fusion process, keeping the remaining three images as ground-truth for evaluation purposes.
Analyzing the results, we can see that the images estimated by the proposed KF-D and SM-D methods produce significantly better visual similarity with the Landsat (ground-truth) images for both bands. For instance, the increase in the island and the expansion of other land parts at date 08/20 are clearly visible for the proposed methods. On the other hand, analyzing the results of the KF-I and SM-I methods, where the temporal dynamics matrix $\bQ_k$ was kept constant and independent of past data, we observe that the results appear very blurry, with a resolution that is comparable to that of the MODIS images. This shows that the proposed weakly supervised calibration strategy is key in order for the KF- and SM-based strategies to obtain high quality reconstructions.




\subsection{Results for larger scale Elephant Butte site}

In this subsection, we compare the proposed strategies to ESTARFM and PSRFM in the Elephant Butte example, which comprises a larger geographical area. For simplicity and to reduce the use of space, we compare only proposed Kalman filter and smoother methods with the block diagonal assumption on the state covariance matrices (i.e., KF-B and SM-B).

\begin{figure*}[h!]
  \centering
  \begin{subfigure}[b]{1\textwidth}
  \includegraphics[width=1\linewidth]{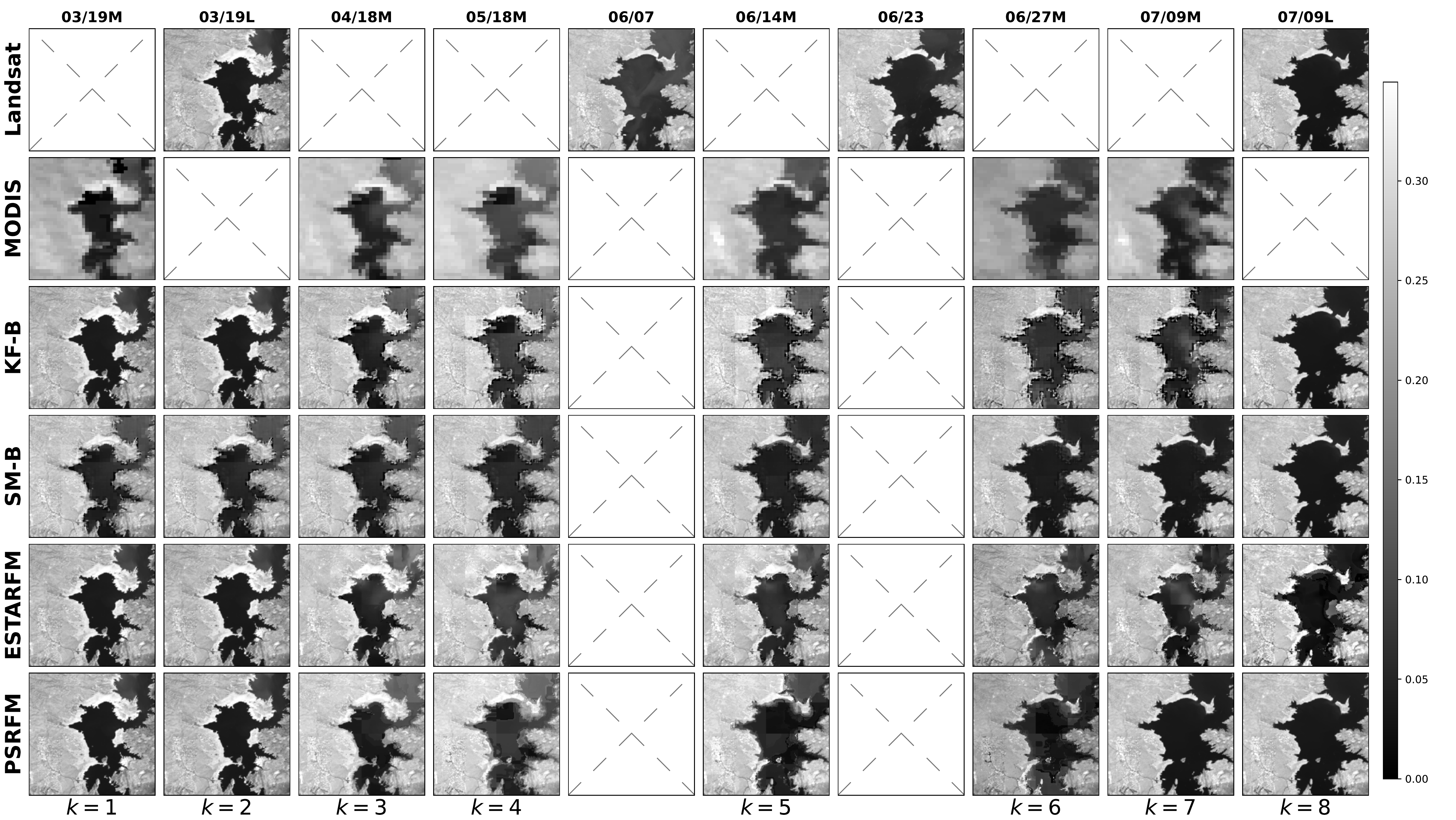}
  \vspace{-0.8cm}
  \caption{Fused images in band 1 (MODIS) and band 4 (LandSat)}
  \label{fig:combination1}
\end{subfigure}
\\
\begin{subfigure}[b]{1\textwidth}
  \includegraphics[width=1\linewidth]{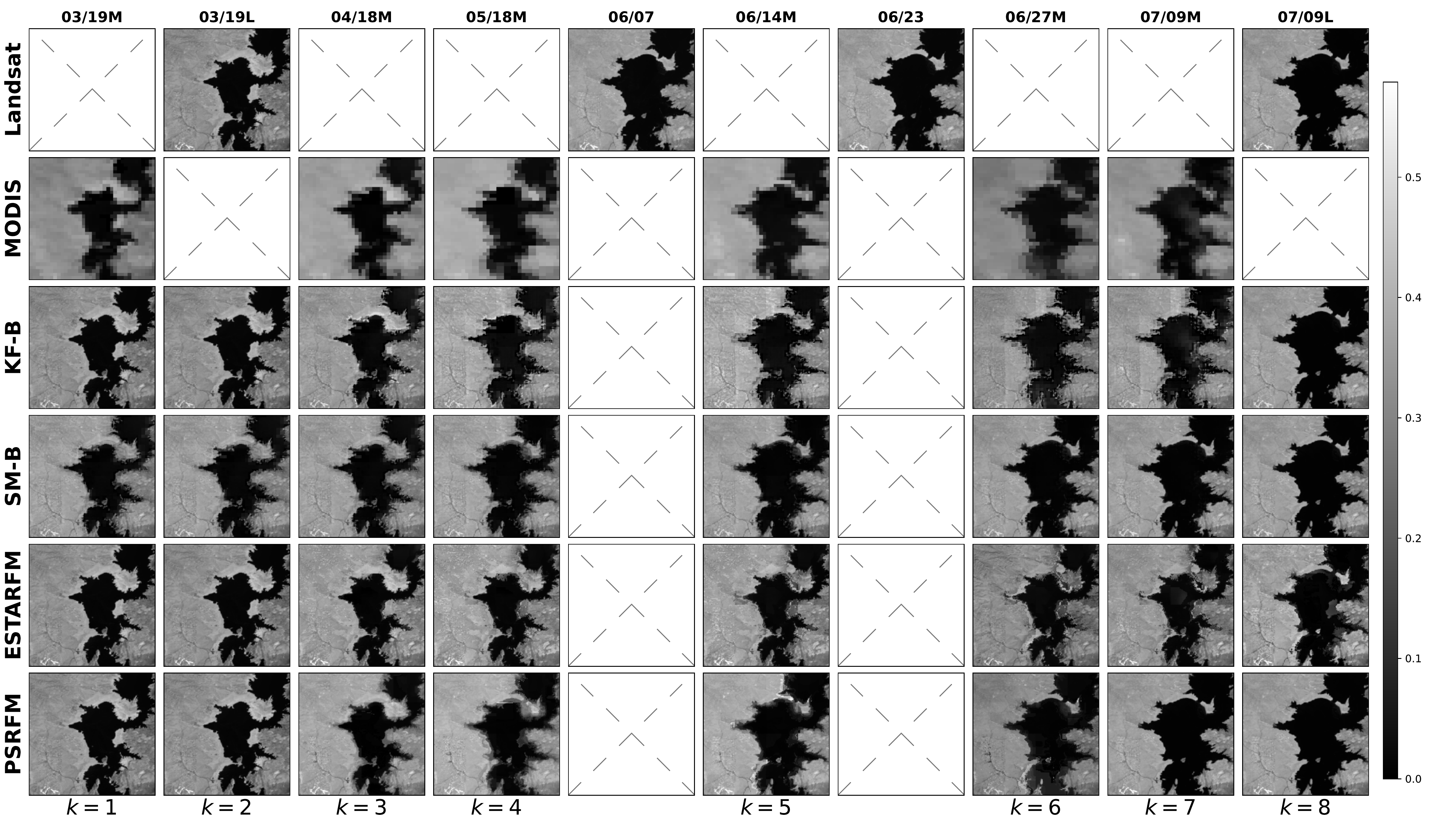}
  \vspace{-0.8cm}
  \caption{Fused images in band 2 (MODIS) and band 5 (LandSat)}
  \label{fig:combination2}
  \end{subfigure}
  \\\vspace{-0.1cm}
  \caption{Fused bands from MODIS and Landsat for the Elephant Butte example using different strategies over time. The first two rows of each subfigure depict MODIS and Landsat bands acquired at dates displayed on top labels. At each time index estimation with KF and SM under block diagonal model assumptions, ESTARFM and PSRFM are presented. Some Landsat images were omitted from the estimation process and used solely as ground-truth. Images used at each update step are indicated on top labels where ``M'' stands for MODIS and ``L'' for Landsat.}\label{fig:combination}
\end{figure*}

\begin{figure}[h]
  \centering
  \includegraphics[width=1\linewidth]{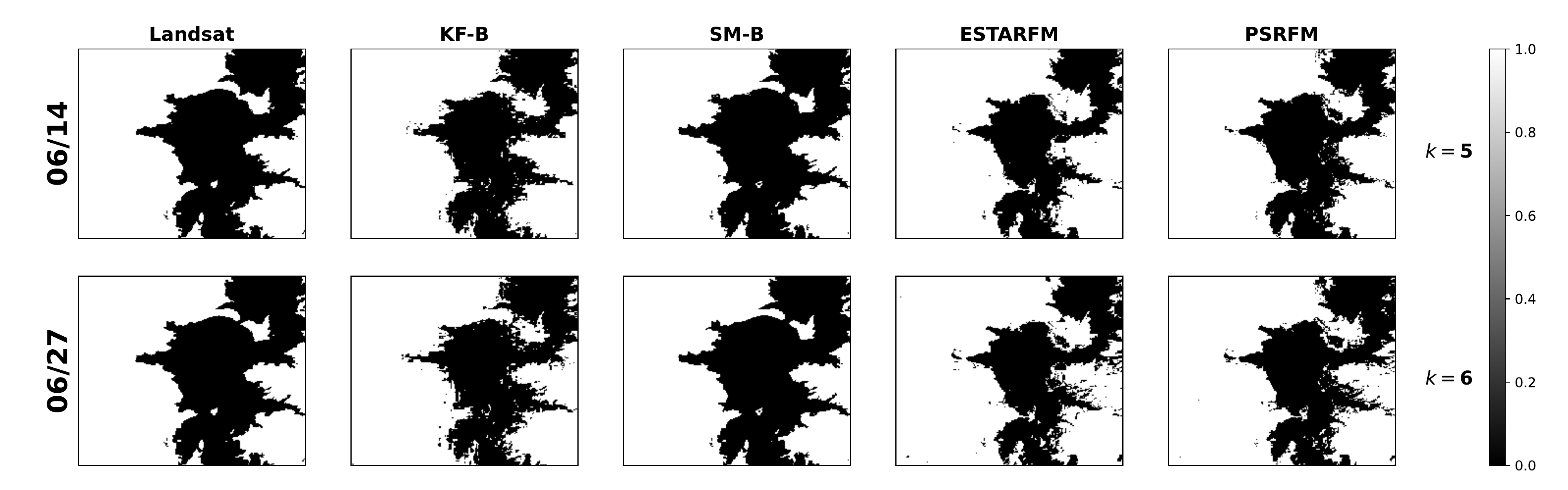}
  \caption{Water map of images for the Elephant Butte example based on K-means clustering strategy where 1 indicates land and 0 indicates water pixels. Unused Landsat classification maps establish the ground-truth (first column).}
  \label{watermap_com}
  \end{figure}

\begin{figure}[h]
  \centering
  \includegraphics[width=1\linewidth]{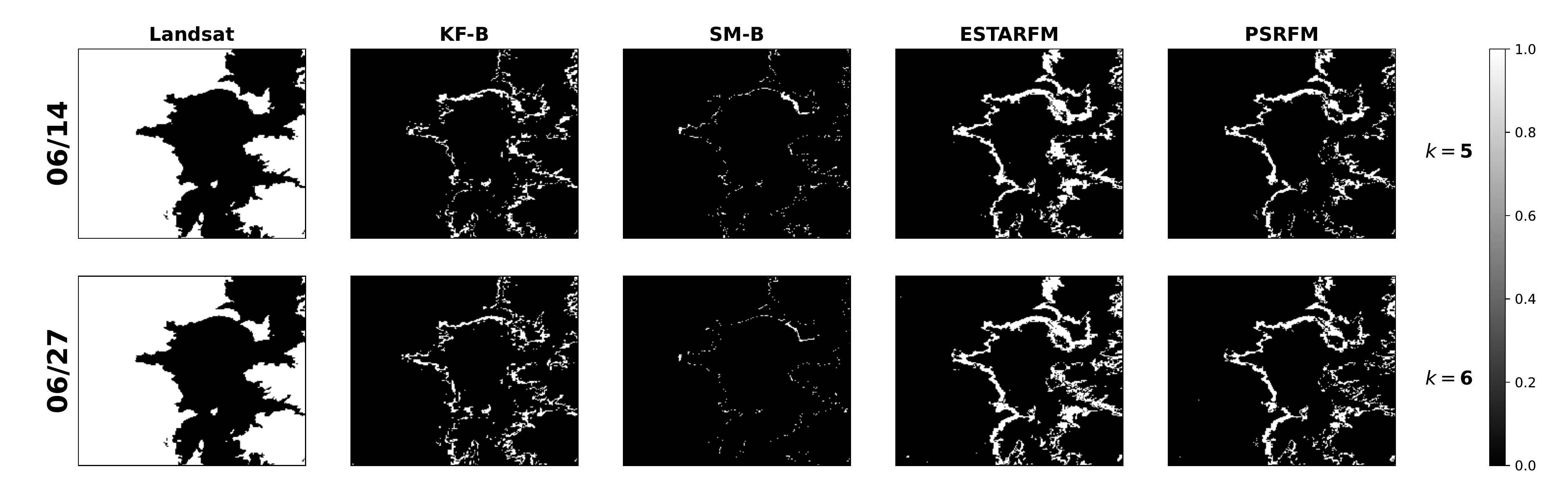}
  \caption{Absolute error of Water map of images for the Elephant Butte example based on K-means clustering strategy. Unused Landsat classification maps establish the ground-truth (first column).}
  \label{watermap_com_error}
  \end{figure}

\begin{figure}[h]
  \centering
  \includegraphics[width=1\linewidth]{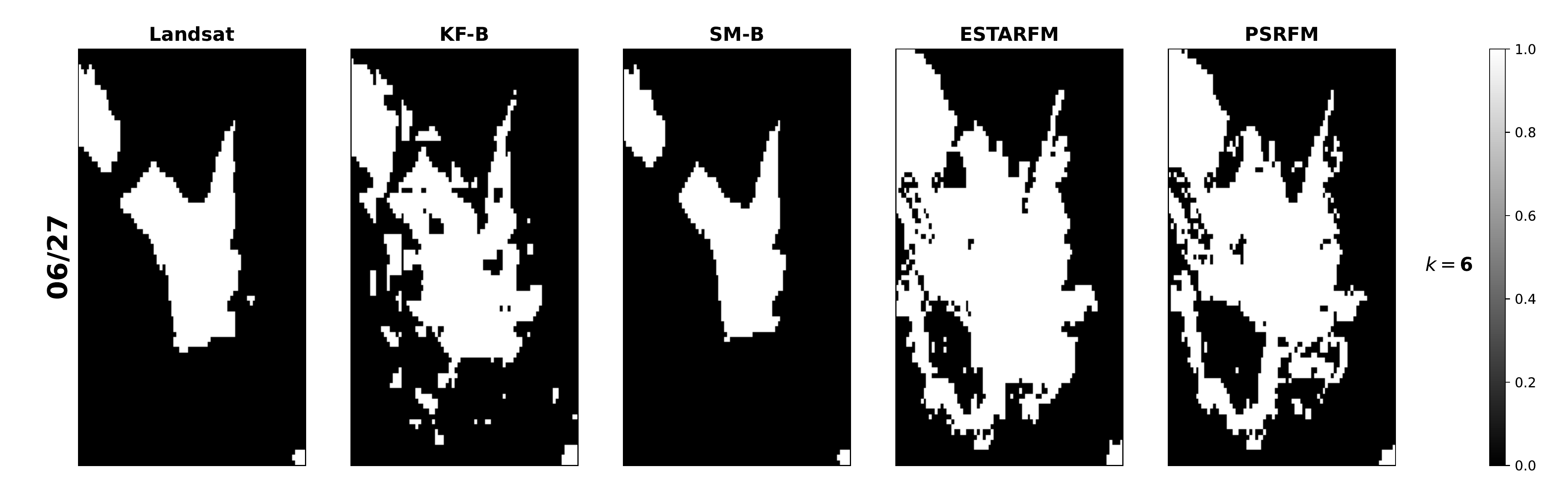}
  \caption{Zoomed-in water map of images for the Elephant Butte example based on K-means clustering strategy where 1 indicates land and 0 indicates water pixels. Unused Landsat classification map at date 06/23 establish the ground-truth (first column).}
  \label{watermap_com_zoomin}
  \end{figure}

\begin{figure}[h]
  \centering
  \includegraphics[width=1\linewidth]{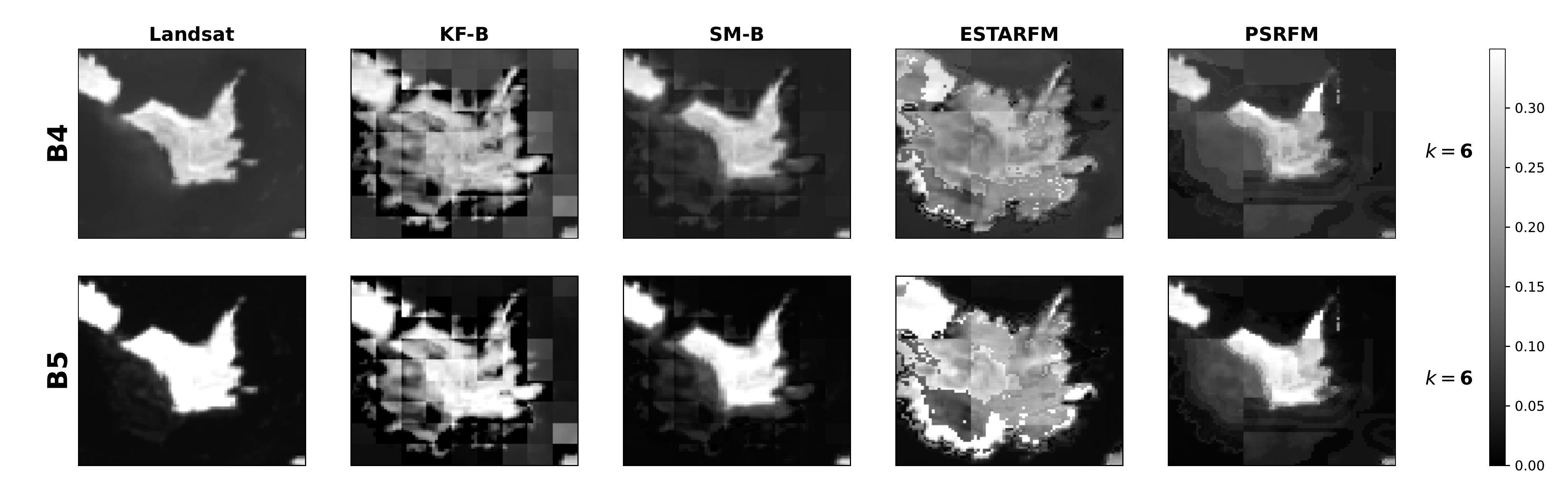}
  \caption{Zoomed-in version of the fused bands from MODIS and Landsat for the Elephant Butte example using different strategies at date 06/27 (ground-truth at 06/23 is shown in the first column).}
  \label{Reconstruction_zoomin}
  \end{figure}

The fusion results for both bands and all algorithms are shown in Figure~\ref{fig:combination}, while Figure~\ref{watermap_com} shows the corresponding water mapping results. To measure the performances of different methods in this large area, the Landsat images at dates 06/07 and 06/23 were chosen as a ground truth to evaluate the quality of the reconstructed images at dates 06/14 and 06/27 (we remark that the MODIS images at dates 06/07 and 06/23 were not available due to the presence of cloud cover). It can be seen that the proposed KF-B, SM-B and the PSRFM methods provide estimates that are close to the ground truth images, whereas the ESTARFM method shows an inferior performance. This can be seen more clearly for the image at date 06/14 ($k=5$), in which the smoother method better captured the increase in the area of the reservoir. To evaluate the performances of different methods more clearly, Figure~\ref{watermap_com_error} shows the absolute error of water maps of images compared with the ground truth, and Figures~~\ref{watermap_com_zoomin} and~\ref{Reconstruction_zoomin} show a zoomed-in area of the image of the fused image and water mapping result, respectively. It can be seen from Figure~\ref{watermap_com_error} that the misclassification errors are concentrated at the borders of the reservoir, which is the area that undergoes the largest amounts of changes over time, and consequently the hardest to classify correctly. The SM-B algorithm shows the best results, followed by KF-B, PSRFM and ESTARFM. Nevertheless, PSRFM provides results that contain less artifacts compared to KF-B, despite the lower classification accuracy. The superior visual quality of the results of SM-B and PSRFM is explained by their use of Landsat images both at the beginning and at the end of the image sequence, whereas KF-B and ESTARFM do not have access to the last Landsat image. 

Table~\ref{tab:SAM_com} presents the SAM results, and Table~\ref{tab:misclassification_com} shows the corresponding percentage of misclassified pixels for the different methods. It can be seen that in terms of SAM, the SM-B method obtained the best results for both dates, followed by PSRFM and ESTARFM. However, the KF-B strategy was able to obtain a better water mapping performance compared to PSRFM. This indicates that the artifacts seen in the (comparatively noisier) reconstructions of KF-B impact the the classification performance in a less substantial way compared to the SAM. This shows that the proposed Kalman-filter based strategy can provide meaningful water mapping results in a real-time setting, in which we do not have access to future Landsat images, precluding smoothing-based algorithms (such as SM-B and PSRFM) to be used.




\begin{table*} [h]
\scriptsize
\centering
\caption{Spectral angle mapper between the estimated high-resolution image and the Land-
sat measurement for the Elephant Butte example (note that the Landsat images at dates 06/07 and 06/23 were not supplied to the algorithms and only used for evaluation purposes).}
\resizebox{0.6\textwidth}{!}{%
\begin{tabular}{c||c|c|c|c}
\hline
Method &  KF-B & SM-B  & ESTARFM & PSRFM \\
\hline\hline
{Image (06/07)} & 5.5416  & \textbf{2.9993}  & {9.2678} & 4.2698 \\
\hline
{Image (06/23)}& 5.7514  & \textbf{1.9923}  & 6.2158 &4.8719 \\
\hline\hline
{Average} &  5.6465 & \textbf{2.4958}  &7.7418 &4.5709
 \\
\hline
\end{tabular}
}
\label{tab:SAM_com}
\end{table*}
\begin{table*} [h]
\scriptsize
\centering
\caption{Percentage of misclassified pixels for the Elephant Butte example (note that the Landsat images at dates 06/07 and 06/23 were not
supplied to the algorithms and only used for evaluation purposes).}
\resizebox{0.6\textwidth}{!}{%
\begin{tabular}{c||c|c|c|c}
\hline
Method &  KF-B & SM-B  & ESTARFM & PSRFM \\
\hline\hline
{Image (06/07)} & 5.3593  & \textbf{1.4289}  & {9.2678} & 6.6606  \\
\hline
{Image (06/23)}& 5.9233  & \textbf{0.8250}  & 10.8330 &{7.8675} \\
\hline\hline
{Average} &  5.6413 & \textbf{ 1.1269}  &10.0504 &{7.2640}
 \\
\hline
\end{tabular}
}
\label{tab:misclassification_com}
\end{table*}




\subsection{Discussion}

The results presented above clearly indicate that the proposed weakly supervised smoother-based image fusion strategy outperforms the ESTARFM and PSRFM algorithms in terms of image reconstruction when an appropriate covariance structure is selected (SM-D and SM-B). This highlights that having less model parameters to estimate (i.e., a more constrained state covariance model) can lead to better results. Moreover, even the Kalman filter strategies (particularly KF-B and KF-D), which estimate high-resolution images from MODIS without having access to any future data, have shown very competitive performance, with great potential for tasks in which high-resolution estimates are required online and one cannot wait for another Landsat image to be available before computing the high-resolution reconstructions.

The advantage of the proposed filter and smoother strategies is more clear when evaluated semantically by means of the water classification performance.
For instance, the growth of the island portion over time in regions that are semantically meaningful leads to more meaningful results that cannot be entirely captured by one standard metric such as the SAM. This can be observed more clearly through the spatial distribution of the misclassification error maps in Figure~\ref{fig:watermap}, which for ESTARFM and PSRFM are significantly more concentrated on the borders between land and water. In general, the proposed filtering-based strategies clearly outperformed both the ESTARFM and PSRFM algorithms, a standard and a state of the art remote sensing image fusion algorithms. Moreover, the proposed distributed implementation, described in Section~\ref{sec:complex}, is able to reduce the computational power and memory demand of the standard Kalman filter and smoother when applied for large images. 



\section{Conclusions}
\label{sec:conclusions}

In this paper, an online Bayesian approach for fusing multi-resolution space-borne multispectral images was proposed.
By formulating the image acquisition process as a linear and Gaussian measurement model, the proposed method leveraged the Kalman filter and smoother to perform image fusion by estimating the latent high resolution image from the different observed modalities. Moreover, a weakly supervised strategy is also proposed to define an informative time-varying dynamical image model by leveraging historical data, which leads to a better localization of changes occurring in the high-resolution image even in intervals where only coarse resolution observations are available.
Experimental results indicate that the proposed strategy can lead to considerable improvements compared to both classical and state-of-the-art image fusion algorithms.

\section{Acknowledgments}
The authors would like to thank the support of the National Geographic Society under Grant NGS-86713T-21, the National Science Foundation under Award ECCS-1845833, and NASA -- GRACE--FO Science Team (80NSSC20K0742).


\clearpage
\bibliographystyle{IEEEtran}
\bibliography{references_MSFus,references_fus1}

\end{document}